\documentclass[]{spie}  

\usepackage{amsmath,amsfonts,amssymb}
\usepackage{graphicx}
\usepackage[colorlinks=true, allcolors=blue]{hyperref}

\usepackage{quantikz}
\usepackage{tcolorbox}
\usepackage{wrapfig}

\newenvironment{mybox}{%
  \begin{tcolorbox}[
    colback=blue!5!white,
    colframe=gray!44!black,
    boxrule=0.5mm, 
  ]%
}{%
  \end{tcolorbox}%
}

\title{ The Gell-Mann feature map of qutrits and its applications in classification tasks}

\author[a]{Themistoklis Valtinos}
\author[a]{Aikaterini Mandilara}
\author[a]{Dimitris Syvridis}
\affil[a]{Department of Informatics and Telecommunications, National and Kapodistrian
University of Athens, Panepistimiopolis, Ilisia, 15784, Greece}

\authorinfo{Further author information:\\T.V.: E-mail: t.valtinos@di.uoa.gr \quad  A.M.: E-mail: mandkat@di.uoa.gr \quad D.S.: E-mail: dsyvridi@di.uoa.gr}

\begin{document} 
\maketitle

\begin{abstract}
Recent advancements in quantum hardware have enabled the realization of high-dimensional quantum states. This work investigates the potential of qutrits in quantum machine learning, leveraging their larger state space for enhanced supervised learning tasks. To that end, the Gell-Mann feature map is introduced which encodes information within an $8$-dimensional Hilbert space. The study focuses on classification problems, comparing  Gell-Mann feature map with maps generated by established qubit and classical models. We test different circuit architectures and  explore possibilities in optimization techniques. By shedding light on the capabilities and limitations of qutrit-based systems, this research aims to advance applications of low-depth quantum circuits.
\end{abstract}


\section{Introduction}
\label{sec:intro}  

Quantum machine learning (QML) \cite{Biamonte_2017, Petru} is an evolving field that seeks to merge the principles of quantum mechanics with machine learning (ML). While research in this domain has predominantly concentrated on qubits, which are the prevalent quantum hardware available at present, this particular work focuses on leveraging the potential of qutrits and putting forth a modular architecture that bears resemblance to a neural network (NN).

The expanded state space offered by qutrits, and qudits in general \cite{Gokhale_2019,Blok_2021,PMID:28658228}, holds significant potential for enhancing ML tasks in a quantum computer\cite{Wash_2023, mandilara2023classification}. These allow for a higher-dimensional encoding of information, and in consequence for the representation of more complex patterns and relationships within the data. With qutrits, the increased granularity of information encoding enables a QML algorithm to capture and process finer details, potentially leading to more accurate and nuanced models.  Quantum algorithms can operate on multiple states simultaneously, leveraging superposition, entanglement, and coherent manipulations. With qutrits, this parallelism can be further amplified, potentially accelerating computations and enabling the exploration of a broader solution space. Finally, the increased state space of qutrits can mitigate the impact of noise and errors during quantum computations. Error correction techniques, such as quantum error correction codes, rely on redundancy within the state space to detect and rectify errors. With a larger state space, qutrits offer more room for implementing robust error correction protocols, enhancing the reliability and stability of QML algorithms.

In the pursuit of enhancing encoding capabilities, in this work we introduce the so called \textit{ Gell-Mann feature map} which draws upon the mathematical framework provided by the special unitary group $SU(3)$. The utilization of the proposed Gell-Mann feature map enables the encoding of information within an $8$-dimensional Hilbert space. This empowers the quantum system to capture and process significantly larger amounts of data even within a single qutrit and introduces a novel avenue for exploration within the realm of QML.

Some prominent designs of quantum circuits used for classification tasks include Quantum Kernels assisted by classical Support Vector Machine methods (QKSVM), Variational Quantum Classifiers (VQCs), and data re-uploading  techniques. These methods differ from each other but share a common component: a quantum circuit that performs a feature map on the input classical data.
In VQCs, the feature map is followed by a variational layer whose parameters can be trained. For data re-uploading, the `unit' composed of the feature map and variational layer is repeated in series, forming a Quantum Neural Network (QNN).
In this study, we adapt VQC, QKSVM, and QNN (formed with data re-uploading method) models to process qutrits and evaluate their performance. Additionally, in some cases, we implement modifications to the original methods, demonstrating advantages. We investigate the performance of our model using several structured datasets and compare it with qubit models and classical ML models.

The structure of this manuscript is as follows: we first review the structure and methods of QKSVM, VQC, and QNN. Next, we provide a basic background on qutrits and introduce the Gell-Mann feature map. We then integrate the Gell-Mann feature map into existing classification methods and address key challenges in each case. Finally, we compare the performance of our optimized designs with established  VQC made of qubits and classical ML methods.

\section{Quantum feature maps and kernels}

The introduction of quantum feature maps and QKSVM can be found in two publications ``Quantum machine learning in feature Hilbert spaces''\cite{Schuld_2019} and ``Supervised learning with quantum-enhanced feature spaces''\cite{Havlicek_2019}. In this work, we closely follow the definitions provided in the former\cite{Schuld_2019}.

A quantum feature map describes the encoding of classical data $\boldsymbol{\vec{x}}$ into a quantum state $|\phi(\boldsymbol{\vec{x}})\rangle$ residing in a Hilbert space $F$. In terms of quantum computation, a quantum feature map $\boldsymbol{\vec{x}}\longrightarrow |\phi(\boldsymbol{\vec{x}})\rangle$ corresponds to a state preparation circuit $\boldsymbol{U}_{\phi}(\boldsymbol {\vec{x}})$ that operates on a ground or vacuum state $|\boldsymbol{0}\rangle=|00..0\rangle$ of the Hilbert space $F$. The unitary operation $\boldsymbol{U}_{\phi}(\boldsymbol {\vec{x}})$ is commonly referred to as the feature-embedding circuit, and it is used to create the state $|\phi(\boldsymbol{\vec{x}})\rangle = \boldsymbol{U}_{\phi}(\boldsymbol {\vec{x}})|\boldsymbol{0}\rangle$.

Suppose that both data points $\boldsymbol{\vec{x}}$ and $\boldsymbol{\vec{x}'}$ have undergone the quantum feature map $\boldsymbol{\vec{x}}\longrightarrow |\phi(\boldsymbol{\vec{x}})\rangle$. The quantum kernel function $K(\boldsymbol{\vec{x}}, \boldsymbol{\vec{x}'})$ measures the similarity between the quantum states corresponding to data points $\boldsymbol{\vec{x}}$ and $\boldsymbol{\vec{x}'}$ and is defined as the modulus squared of their inner product:
\begin{equation}
    K(\boldsymbol{\vec{x},\vec{x}'}) = |\langle\phi(\boldsymbol{\vec{x}})|\phi(\boldsymbol{\vec{x}'})\rangle|^2~~. \label{Ker}
\end{equation}

\begin{figure}[ht]
\begin{mybox}
\begin{center}
\begin{quantikz}
\lstick{\ket{\textbf{0}}} &\gate{\boldsymbol{U}_{\phi}(\boldsymbol {\vec{x}})}&  \gate{\boldsymbol{U^\dagger}_{\phi}(\boldsymbol {\vec{x}'})} &\meter{}
\end{quantikz}
\end{center}
\caption{ \label{fig:1qutkern} A quantum circuit for  estimating the kernel in Eq.(\ref{Ker}).}
\end{mybox}
\end{figure}
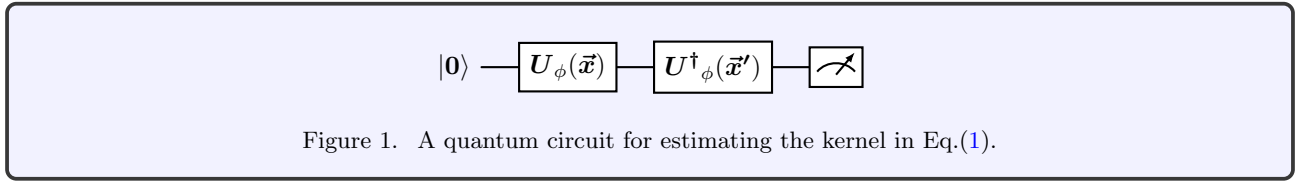

To implement the kernel, we have the option to prepare two states, \(|\phi(\boldsymbol{\vec{x}})\rangle\) and \(|\phi(\boldsymbol{\vec{x}'})\rangle\), on separate sets of qubits and measure their overlap using a SWAP test routine. Alternatively, in the implicit approach \cite{Schuld_2019} depicted in Figure~\ref{fig:1qutkern}, a quantum computer implements the mapping $\boldsymbol{\vec{x}}\longrightarrow |\phi(\boldsymbol{\vec{x}})\rangle$ via a feature-embedding circuit $\boldsymbol{U}_{\phi}(\boldsymbol {\vec{x}})$, subsequently apply the inverse encoding $\boldsymbol{U^{\dagger}}_{\phi}(\boldsymbol {\vec{x}'})$ with the second data point and estimate the inner product. 

 After the estimation of the kernel via the quantum circuits, the subsequent training algorithm is delegated to a classical device which employs the kernel as input for a Support Vector Classifier (SVC). For brevity in this work, we refer to this approach which involves estimating the kernel $K(\boldsymbol{\vec{x},\vec{x}'})$ via a quantum circuit and performing classification using Support Vector Machine methods as QKSVM.

\section{VQCs and a model for QNN}

\begin{figure}[ht]
  \centering
  \includegraphics[width=5.3in]{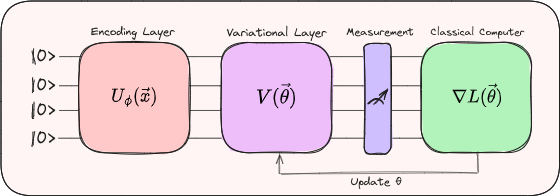}
  \caption{\label{Fig2} The basic structure of a VQC}
\end{figure}

A VQC is a hybrid quantum-classical optimization algorithm in which an objective function is evaluated via a variational quantum circuit, and the parameters of this function are updated using classical optimization methods \cite{Havlicek_2019,Schuld_2019,Schuld_2020}. 
 In Fig.~\ref{Fig2}  we schematically present the three main `components'  of a VQC:
\begin{enumerate}
  \item Feature Map. This consists of the feature-embedding circuit that performs the unitary transformation $\boldsymbol{U_\phi(\vec{x})}$ mapping the classical input data $\boldsymbol{\vec{x}}$ to a quantum state $ |\phi(\boldsymbol{\vec{x}})\rangle$.
  
  \item Variational Circuit. The variational quantum circuit contains gates with adjustable parameters $\boldsymbol{\vec{\theta}}$ and can be expressed as a unitary transformation $\boldsymbol{V(\vec{\theta})}$ that depends this set of trainable parameters. The output state of the variational layer can be then represented as: $\boldsymbol{|\psi_{out}\rangle} = \boldsymbol{V(\vec{\theta}) U_\phi(\vec{x}) |\psi_{in}\rangle}$, where $\boldsymbol{|\psi_{in}\rangle}$ is the initial state of the quantum system. Usually $\boldsymbol{|\psi_{in}\rangle}=|\boldsymbol{0}\rangle$.
  
  \item Measurement. At the end of the quantum circuit an observable $\boldsymbol{O}$ is measured and  its mean value is interpreted as the output of the classifier. The measurement process can be represented as a function of the output state: $\text{label}(\boldsymbol{\vec{x}}) = \langle\boldsymbol{\psi_{out}}|\boldsymbol{O}|\boldsymbol{\psi_{out}}\rangle$.
\end{enumerate}

In the training phase, the objective is to determine the optimal values for $\boldsymbol{\vec{\theta}}$ that yield the most accurate predictions. To achieve this, the classifier employs classical optimization algorithms. It compares the predicted labels $\boldsymbol{\hat{y}}$ with the actual labels $\boldsymbol{y}$ provided in the training data and computes the loss using a cost function $\boldsymbol{L(\vec{\theta})}$. The classifier then iteratively adjusts the parameters of the variational circuit based on the computed loss. This iterative process continues until the cost function reaches a stable state \cite{Schuld_2020}.

\subsection{A model of QNN employing data re-uploading}
\begin{figure}[ht]
  \centering
  \includegraphics[width=5.34in]{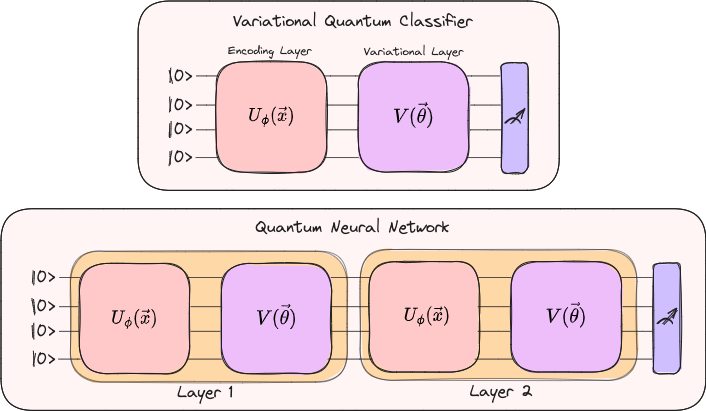}
  \caption{Example of a QNN structure consisting of two VQC layers arranged in a sequential stacking configuration.}
\end{figure}

While there exist various architectures of QNNs that have been explored in the literature, in this work we adopt a straightforward approach of using a stack of VQC layers. A VQC consisting of an encoding and variational layer can be seen as an analog of a classical perceptron. It is thus natural to create an analog of multilayered NNs by stacking two or more VQC circuits in series. This architectural concept, was firstly introduced in ``Data re-uploading for a universal quantum classifier''\cite{P_rez_Salinas_2020}, and it results in highly nonlinear mappings (since the data are re-uploaded multiple times). Additionally, the number of trainable parameters increases linearly with the number of VQC layers.

 The depth of this QNN model refers to the number of stacked VQC layers. In practical implementations, we have observed that more complex problems often require a deeper QNN to capture intricate features and relationships within the data. Increasing the depth allows for a more expressive representation of the problem, enabling the QNN to learn and classify complex patterns. 
However, the choice of QNN depth, as for NNs, involves a trade-off between computational resources and performance. While deeper QNNs may offer better representation capabilities, these demand longer gate sequences, and increased optimization complexity. Therefore, determining the appropriate depth for a QNN should be a careful consideration based on the available quantum hardware and the specific requirements of the problem at hand.

\subsection{Optimization}

In the context of ML, optimizers are algorithms used to iteratively update the parameters of a model in order to minimize the loss function. Different optimizers use different strategies for updating the parameters, with some methods being more efficient or effective than others depending on the problem at hand\cite{Schuld_2019_grad}.

Gradient-based methods aim to identify an optimal solution by finding a point where the gradient is equal to zero. Some common optimizers include stochastic gradient descent (SGD), which updates the parameters in the direction of the negative gradient of the loss function, and variants such as momentum-based methods, which use a moving average of past gradients to smooth the update process and accelerate convergence. For the purposes of this paper, optimizers from the SGD family were primarily used, specifically Adam and RMSProp proposed by Geoffrey Hinton, with the latter performing better.

RMSprop, short for Root Mean Square Propagation,  adapts the learning rate for each parameter individually based on the magnitude of recent gradients. This adaptive learning rate helps to mitigate the issues of vanishing or exploding gradients, making it suitable for training deep neural networks. The algorithm maintains a moving average of the squared gradients for each parameter, which is then used to update the parameters. By dividing the learning rate by the square root of this average, RMSprop scales the learning rate based on the history of gradients. This helps to converge faster in regions with consistent gradients while being more cautious in fluctuating or noisy regions.

In more details,  in RMSProp  the learning rate $\eta$ for updating a weight $w$
\begin{equation}
w := w - \frac{\eta}{\sqrt{v(w,t)}} \nabla Q_{i}(w)
\end{equation}
 is divided by  by a running average $v(w,t)$ of the magnitudes of recent gradients for that weight. With $Q_{i}(w)$ we denote the loss function for the $i$th train data point.  The running average is calculated in terms of mean square as:
\begin{equation}
v(w,t) := \gamma v(w,t-1) + (1-\gamma) \left(\nabla Q_{i}(w)\right)^{2}
\end{equation}
The concept of storing the gradients as a sum of squares is borrowed from Adagrad and the ``forgetting'' via the factor $\gamma$, is introduced to solve Adagrad's diminishing learning rates in non-convex problems by gradually decreasing the influence of old data \cite{hinton_2012_lecture}. 

RMSProp has shown good adaptation of learning rates and is capable of working with mini-batches. Still, gradient-based optimization is challenging as it may converge slowly and can fail to find the global optimum due to the presence of multiple local minima in the optimization problem. In these scenarios, gradient-free methods can be a useful alternative, as they can overcome the issue of local minima. However, these require higher computational capacity, especially for problems with high-dimensional search spaces which one encounters with variational quantum computing.

\section{ Qutrits and the Gell-Mann feature map}

Qubits represent quantum systems that yield two mutually exclusive outcomes when an observable is measured, such as energy. Their underlying symmetry group is the $SU(2)$ group and its generators, the Pauli matrices, form the basis for constructing quantum gates and related operations.
Qutrits extend the concept of a qubit to the case of three mutually exclusive measurement outcomes and their symmetry group is the    $SU(3)$ group. The Gell-Mann matrices,
\begin{align*} 
\boldsymbol{\lambda}_1 &= \begin{pmatrix} 0 & 1 & 0 \\ 1 & 0 & 0 \\ 0 & 0 & 0 \end{pmatrix} &
\boldsymbol{\lambda}_2 &= \begin{pmatrix} 0 & -i & 0 \\ i & 0 & 0 \\ 0 & 0 & 0 \end{pmatrix} &
\boldsymbol{\lambda}_3 &= \begin{pmatrix} 1 & 0 & 0 \\ 0 & -1 & 0 \\ 0 & 0 & 0 \end{pmatrix} &
\boldsymbol{\lambda}_4 &= \begin{pmatrix} 0 & 0 & 1 \\ 0 & 0 & 0 \\ 1 & 0 & 0 \end{pmatrix} \\
\boldsymbol{\lambda}_5 &= \begin{pmatrix} 0 & 0 & -i \\ 0 & 0 & 0 \\ i & 0 & 0 \end{pmatrix} &
\boldsymbol{\lambda}_6 &= \begin{pmatrix} 0 & 0 & 0 \\ 0 & 0 & 1 \\ 0 & 1 & 0 \end{pmatrix} &
\boldsymbol{\lambda}_7 &= \begin{pmatrix} 0 & 0 & 0 \\ 0 & 0 & -i \\ 0 & i & 0 \end{pmatrix} &
\boldsymbol{\lambda}_8 &= \begin{pmatrix} \frac{1}{\sqrt{3}} & 0 & 0 \\ 0 & \frac{1}{\sqrt{3}} & 0 \\ 0 & 0 & \frac{-2}{\sqrt{3}} \end{pmatrix}
\end{align*}
 comprise a set of eight linearly independent $3\times3$ traceless Hermitian matrices in the computational basis of a qutrit and  span the Lie algebra of the $SU(3)$ group. 
 Gell-Mann matrices satisfy the orthogonality condition $\text{Tr}(\boldsymbol{\lambda}_i\boldsymbol{\lambda}_j) = 2\delta_{ij}$ and can be utilized as generators to construct elements of the $SU(3)$ group in the same way Pauli matrices are used for the $SU(2)$ group operations\cite{etingof2022lie}. Consequently, any unitary operation $\boldsymbol{U}$ on a qutrit can be expressed (up to a phase) as follows:
\[
\boldsymbol{U} = e^{i\sum_a\theta^a \boldsymbol{\lambda}_a},
\]
where $\theta^a$ are real parameters, and $\boldsymbol{\lambda}_a$ represents the Gell-Mann matrices\cite{merker2010theory}.

In the case of qubits, various feature maps can be employed to encode classical data into quantum states. The choice of a feature map depends on the specific problem and the desired data representation. In practice, rotation encoding is often the preferred option. One such rotation encoding feature map is the Pauli Feature Map that uses the Pauli matrices to encode classical data into a quantum state. In analogy, we introduce the Gell-Mann feature map, which is also a rotation encoding feature map defined as:
\begin{equation}
\label{eq:gelrot}
\boldsymbol{\operatorname{R}}(w_1,w_2,\dots,w_8) = \exp[-i\sum_{a=1}^8w_a \boldsymbol{\lambda}_a],
\end{equation}
where $\boldsymbol{\lambda}_a$ represents the $a$th Gell-Mann matrix, and $w_a$ can be restricted in $\left[0,2\pi\right]$.  

The proposed Gell-Mann feature map allows us to encode information within an 8-dimensional Hilbert space, significantly enhancing the quantum system's data processing capabilities within a single qutrit. Notably, this can be achieved without the need for entangling operations.

It is essential to clarify that in this study, the Gell-Mann feature map does not refer to a specific arrangement of gates, but rather to the use of Gell-Mann rotation operators in the encoding process. In other words, not all elements need to be present in the summation in Eq.~(\ref{eq:gelrot}) and in the special case where only one generator $\boldsymbol{\lambda}_a$ is present, in what follows, we denote the rotation as $\boldsymbol{R}_{\boldsymbol{\lambda}_a}(w)=\exp[-i w \boldsymbol{\lambda}_a]$.

\subsection{Realization of qutrits in practice}

Qutrits, and more broadly, qudits, offer a promising avenue for achieving significant improvements in quantum circuit decomposition and cost reductions for essential quantum algorithms, paving the way for scalable quantum computation. Notably, optical quantum states based on entangled photons lie at the core of quantum information science, serving as pivotal building blocks for solving questions in fundamental physics and are at the heart of quantum information science. Integrated photonics has become a leading platform for the compact, cost-efficient, and stable generation and processing of non-classical optical states. 

The paper ``In-chip Generation of High-Dimensional Entangled Quantum States and Their Coherent Control'' by Michael Kues et al explores the generation, manipulation, and control of high-dimensional entangled quantum states on a chip \cite{PMID:28658228}. The authors highlight the advantages of using qudits, over conventional qubit-based systems in terms of increased information capacity and improved resilience against noise and errors.
The paper describes the experimental implementation of high-dimensional quantum states using integrated photonics. The authors discuss the use of integrated waveguide circuits on a chip to generate and manipulate entangled photon pairs with high-dimensional quantum states. They also delve into the coherent control of high-dimensional entangled states on the chip, through the utilization of reconfigurable waveguide circuits for precise manipulation of photon states. 
Additionally, the paper addresses the characterization and measurement of high-dimensional entangled states. The implementation of state tomography techniques is discussed, enabling accurate determination of the quantum states produced on the chip. The measurement results validate the successful generation of high-dimensional entanglement and demonstrate the functionality of the on-chip platform to perform deterministic high-dimensional gate operations.

Another notable approach can be found in the work ``Asymptotic Improvements to Quantum Circuits via Qutrits''\cite{Gokhale_2019} where the authors focus on the Generalized Toffoli gate, an important primitive in quantum algorithms. They present a construction that uses qutrits and doesn't require ancilla bits, which are extra bits used to implement irreversible logical operations in quantum computing. Instead, the qutrit's third state is used to store temporary information.
The Generalized Toffoli gate, has been studied before, but previous circuit constructions for this gate, such as the Gidney, He, and Barenco designs, rely on qubits and have different tradeoffs in terms of circuit depth and ancilla usage. In \cite{Gokhale_2019} the qutrit-based construction, eliminates the need for ancilla bits by directly storing temporary information in the qutrit controls. This eliminates the requirement for ancilla bits and improves the efficiency of the circuit. This new construction results in significant improvements in circuit depth and gate count for important quantum algorithms like Grover's search and Shor's factoring algorithm. It also offers a more favorable trade-off between information compression and higher per-qudit errors, justifying the use of qutrits in quantum computing.

Quantum computing company Rigetti is exploring experimental new hardware configurations that could improve the performance of its quantum processors by introducing a third energy state to its qubits, thus turning them into qutrits. Their superconducting quantum processors are based on the transmon design \cite{PhysRevA.76.042319}, which allows them to access multiple energy states and offers the advantages of increased information encoding and decreased readout errors. One key factor contributing to these advantages is the significantly larger state space accessible using qutrits compared to qubits. Single qutrit operations reside in $SU(3)$, while two-qutrit operations exist in $SU(9)$, resulting in a more than twofold increase in dimensionality compared to the two-qubit case.

Finally, another notable paper ``Quantum Information Scrambling on a Superconducting Qutrit Processor'' \cite{Blok_2021} explores the dynamics of quantum information in strongly interacting systems, known as quantum information scrambling. This phenomenon has recently become a common thread in understanding black holes, transport in exotic non-Fermi liquids, and many-body analogs of quantum chaos. Previously, verified experimental implementations of scrambling focused on systems composed of two-level qubits.
The authors take the first steps toward accessing such phenomena by realizing a quantum processor based on superconducting qutrits. They demonstrate the implementation of universal two-qutrit scrambling operations and embed them in a five-qutrit quantum teleportation protocol. The measured teleportation fidelities (Favg=0.568±0.001) confirm the presence of scrambling even in the presence of experimental imperfections and decoherence \cite{Blok_2021}.
The development of a superconducting five-qutrit processor that runs a quantum teleportation algorithm serves as a proof of principle. The authors engineer two new ways of entangling superconducting qutrits, placing them in a non-classical state required for quantum computing. While it is still too early for these studies to teach us something new about quantum information propagation, similar algorithms run on future quantum processors might shed light on such fundamental questions.

\section{QKSVM with qutrits}

\subsection{Architecture}

\begin{figure}[ht]
\begin{mybox}
\begin{center}
\begin{quantikz}
\lstick{\ket{\textbf{0}}} &\gate{\textbf{H}}&\gate{\boldsymbol{R_{\lambda_1}}(x_{1})}&\gate{\boldsymbol{R_{\lambda_2}}(x_{2})}&\gate{\boldsymbol{R_{\lambda_3}}(x_{3})}&\gate{\boldsymbol{R_{\lambda_4}}(x_{4})} {}
\end{quantikz}
\end{center}
 \caption{ \label{fig:1qutfmap} A single qutrit feature map, where four features are encoded using the first four Gell-Mann matrices.}
  \end{mybox}
\end{figure}
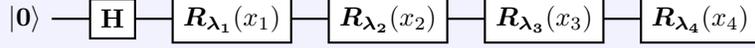

In the process of obtaining the results of this work, different combinations of rotation gates were explored for encoding. While some datasets required alternative gate arrangements, for the majority of datasets, the following two kernel architectures proved the most effective. The first and simplest kernel involved using a single qutrit as demonstrated in Fig.~\ref{fig:1qutfmap}.

Providing more details on the scheme of  Fig.~\ref{fig:1qutfmap}. There, one first applies a
 generalized Hadamard gate, also known as the quantum Fourier transform (QFT) gate, to produce a superposition of basis states of the qutrit. This is an extension of the Hadamard gate on a qubit, represented by the matrix
    \[
    \textbf{H} = \frac{1}{\sqrt{3}}
    \begin{bmatrix}
    1 & 1 & 1 \\
    1 & \omega & \omega^2 \\
    1 & \omega^2 & \omega
    \end{bmatrix}
    \]
where $\omega = e^{(2\pi i/3)}$.
 Subsequently, the features are encoded onto the qutrit using the first four Gell-Mann matrices. 

The second data point $\boldsymbol{\vec{x}'}$ is encoded using the conjugate transpose of these gates, symbolized with $\boldsymbol{U^\dagger(\vec{x}')}$ on this and subsequent circuits. The resulting states are then used to compute the quantum kernel by taking the inner product between them as demonstrated in Fig.~\ref{fig:1qutkern}. This quantum kernel can then be passed to a classical SVM classifier.

\begin{figure}[ht]
\begin{mybox}
\begin{center}

\begin{quantikz}
\lstick{\ket{\textbf{0}}} &\gate{\textbf{H}}&\gate{\boldsymbol{R_{\lambda_1}}(x_{1})}&\gate{\boldsymbol{R_{\lambda_2}}(x_{2})}&\gate{\boldsymbol{R_{\lambda_3}}(x_{3})}&\gate{\boldsymbol{R_{\lambda_4}}(x_{4})}&\gate[2]{\boldsymbol{LZZ}} \\
\lstick{\ket{\textbf{0}}} &\gate{\textbf{H}}&\gate{\boldsymbol{R_{\lambda_5}}(x_{1})}&\gate{\boldsymbol{R_{\lambda_6}}(x_{2})}&\gate{\boldsymbol{R_{\lambda_7}}(x_{3})}&\gate{\boldsymbol{R_{\lambda_8}}(x_{4})} &
\end{quantikz}
\end{center}
 \caption{ \label{fig:2qutfmap} A two qutrit feature map, where four features are encoded twice using all eight Gell-Mann matrices. Subsequently, the entangling operator LZZ is applied to the two qutrits.}
  \end{mybox}
\end{figure}
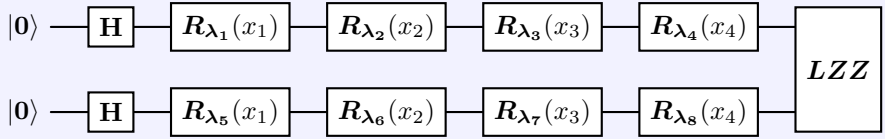

The second circuit exhibiting  good performance  during simulations is shown in Fig.~\ref{fig:2qutfmap}. This involves  two qutrits with each one being placed in a superposition state using a Hadamard operator. Then the features are encoded on the first qutrit using the first four Gell-Mann matrices, and on the second qutrit using the last four Gell-Mann matrices. Subsequently, an entangling operation is performed between them by applying the $\boldsymbol{LZZ}$ gate, defined as follows:
\begin{equation}
\boldsymbol{LZZ} = \exp\left [-i \boldsymbol{LZ} \otimes \boldsymbol{LZ}\right ]
\end{equation}
where the single-qutrit operator  $\boldsymbol{LZ}=\boldsymbol{\lambda}_3 + \sqrt{3}\boldsymbol{\lambda}_8$.

The circuits of Figs.~\ref{fig:1qutfmap}-\ref{fig:2qutfmap} illustrate the encoding process of the four features of the Iris dataset. When datasets contain more than four features, stacked layers are employed  to accommodate the additional features.

\subsection{Results}

A quantum kernel alone cannot be used to make predictions on a dataset, but only serves as a tool to measure the overlap between two data points. For the SVM, sci-kit learn's SVC is employed, which requires a kernel function that takes as input two sets of data points. To validate each kernel, a check was performed to ensure that evaluating the kernel of a data point with itself returned 1, since the simulation is noiseless. 

In the majority of cases investigated, and if not differently stated, we employ two qutrits with the kernel architecture of Fig.~\ref{fig:2qutfmap} that yields the best results and effectively captures the complexity of the data sets. To evaluate the performance of the classifier, the percentage of correctly classified data points in the dataset is measured.

\subsubsection{Binary Classification}

We start our tests with  binary classification tasks and two-dimensional problems, specifically Circles, XOR, and Moons, for being able to visualize the achieved separation capacity. We employ the feature map of Fig.~\ref{fig:2qutfmap} with   the two features (coordinates) uploaded twice so that all Gell-Mann rotations are in use in this model.


\begin{wrapfigure}{l}{0.49\textwidth}
  \centering
  \includegraphics[height=7.2cm]{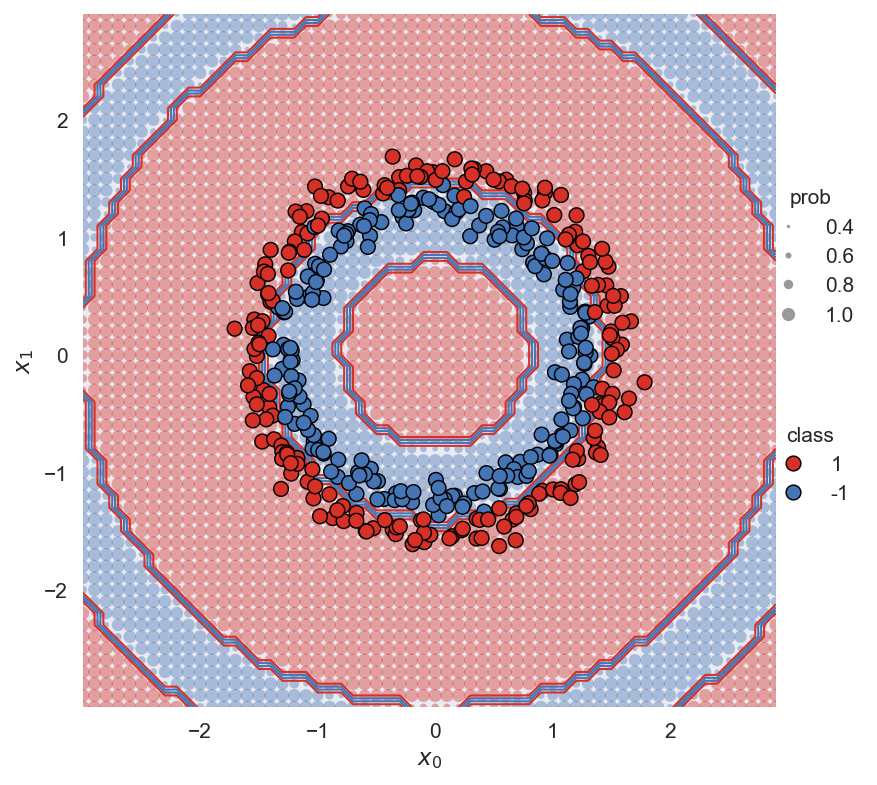}
  \caption{ \label{fig:qksvm1} Decision Boundaries of QKSVM with two qutrits. Circles Accuracy: $100\%$}
\end{wrapfigure}

\begin{figure} [ht]
\begin{center}
\begin{tabular}{c} 
\includegraphics[height=7.7cm]{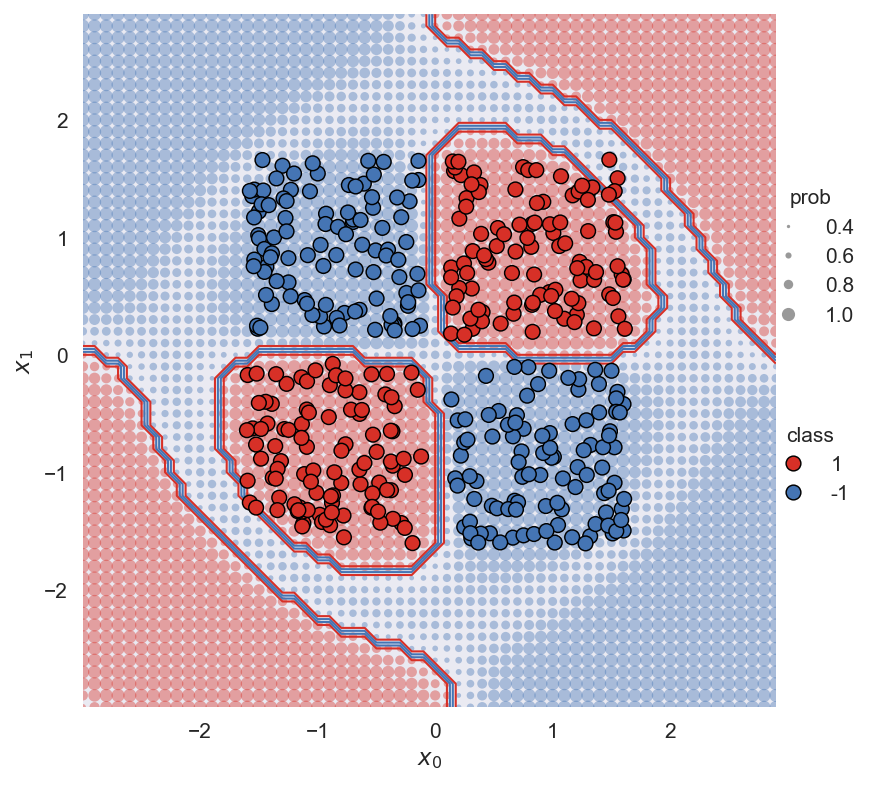}
\includegraphics[height=7.7cm]{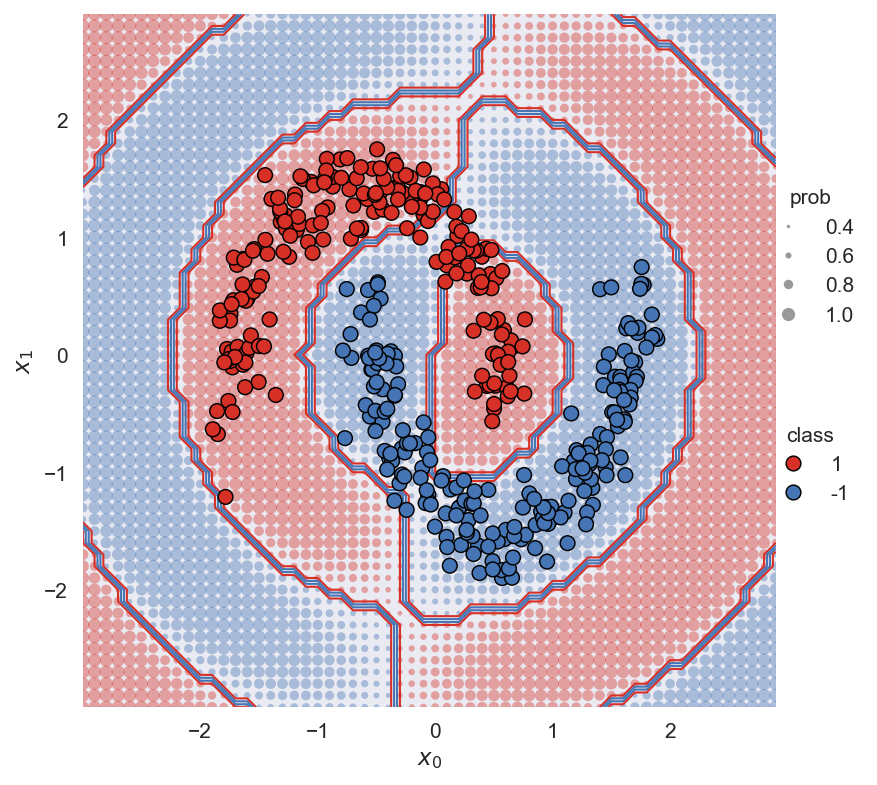}
\end{tabular}
\end{center}
\caption{ \label{fig:qksvm2} Decision Boundaries of QKSVM with two qutrits. XOR Accuracy: $99.17\%$. MOONS Accuracy: $96.67\%$.}
\end{figure}

A decision boundary is a hyper-surface that partitions the underlying vector space into one set for each class. To gain insight into the model's fitting process, the decision boundaries for the three tasks  are plotted in Figs.~\ref{fig:qksvm1}-\ref{fig:qksvm2} using the data visualization functions from Tim von Hahn's blog \cite{beautifulplots}.
In addition to illustrating the decision boundaries between classes, these plots provide a visual representation of the probability assigned to a data point belonging to a specific class. This is achieved by adjusting the size of the dots, with larger dots indicating a higher likelihood of the data point being classified into that particular class.
This probability of a prediction is often referred to as confidence, and it is distinct from accuracy. Confidence reflects the model's certainty in its prediction, while accuracy measures the overall correctness of the model's predictions across a dataset. 

In Fig.~\ref{fig:qksvm1}, one can observe that  the quantum kernel combined with the SVM achieves an accuracy of 100\% on the Circles dataset, showcasing its ability to accurately classify all instances in this dataset.


The benchmark results, presented in in Fig.~\ref{fig:qksvm2}, demonstrate that the model achieves also a high accuracy of 99.17\% on the XOR dataset, and an accuracy of 96.67\% on the Moons dataset. It is important to note that the accuracy was evaluated on the test set, while the decision boundaries were on the entire dataset.

The decision boundaries of all three binary problems indicate that the outer points in the datasets can be accurately classified. The majority of the data instances fall within the correct class for circles, and for the other two slightly more complex datasets, the performance remains nearly perfect. Furthermore, no strong artifacts that would raise concerns about the model's reliability are observed.

\subsubsection{Multiclass Classification}

For the multiclass classification tasks, we test datasets with four or higher input dimensions. and stacked layers were utilized for the encoding. This approach involved breaking down the datasets and using multiple layers of encoding for each set of features. The number of different classes was set to three.

The results in Table~\ref{tab:qksvmaccuracy} are obtained via the two-qutrit kernel presented in Fig.~\ref{fig:2qutfmap}, except for the Wine dataset where Hadamard gates were not utilized. Since the SVM method relies on scikit-learn's SVC, the decision boundaries and support vectors are determined by the input data. Therefore, conducting multiple benchmarks is unnecessary, since running it with identical inputs consistently yields the same outcomes.

\begin{table}[ht]
    \caption{Metrics of the performance of the two-qutrit QKSVM on multi-class classification tasks.}
    \label{tab:qksvmaccuracy}
    \begin{center}
    \begin{tabular}{|l|l|l|l|l|}
    \hline
                       & \textbf{IRIS} & \textbf{WINE} & \textbf{GLASS} & \textbf{SEED} \\ \hline
    \textbf{Recall}    & 90.00\%       & 83.97\%       & 90.48\%       & 88.10\%       \\ \hline
    \textbf{Precision} & 92.31\%       & 83.51\%       & 90.43\%       & 89.56\%       \\ \hline
    \textbf{Accuracy}  & 90.00\%       & 83.33\%       & 90.24\%       & 88.10\%       \\ \hline
    \textbf{F1-score}  & 89.77\%       & 83.42\%       & 90.21\%       & 87.93\%       \\ \hline
    \end{tabular}
    \end{center}
\end{table}

The model achieved high recall scores across all datasets, ranging from 83.97\% to 90.48\%. This indicates that the model successfully identifies a significant portion of true positive instances within each class. The precision scores vary, with values between 83.51\% and 92.31\%, which shows that the model's ability to accurately classify instances within each predicted positive class varies across the multiclass datasets. 
The accuracy scores range from 83.33\% to 90.24\%, achieving an overall good level of correctness in its predictions across the majority of datasets. The F1-scores, which provides a balanced evaluation of the model's performance, range from 83.42\% to 89.77\% and reflect its ability to balance precision and recall for each class. 

The obtained metrics demonstrate that the quantum kernel effectively maps the features in the qutrits, enabling the SVM to accurately predict the target class. These findings provide a solid foundation for assessing the efficacy of the Gell-Mann feature map prior to its integration within the QNN framework, which incorporates a variational layer. This transition inherently directs attention towards the optimization process of the parameter as the central challenge. 

\section{QNN with qutrits}

For the quantum kernel a hybrid quantum-classical method was employed with the help of NumPy and scikit-learn, but in the case of  QNN, with the introduction of the variational layer and the increase in the number of operations involved, PyTorch was chosen. PyTorch facilitates the explicit definition of the model's architecture and is efficient in handling complex operations. 

The QNN class is implemented as a module that encapsulates the functionality of the model and inherits from PyTorch's \texttt{Module} class. This design choice allows for smooth integration with PyTorch's optimization algorithms and cost functions. After evaluating various optimizers, \texttt{RMSprop} was chosen for training. Additionally, considering that the datasets are multiclass, the \texttt{cross-entropy} loss function was selected, making the model well-equipped to handle binary and multiclass problems effectively.

To address the challenge of barren plateaus \cite{Barren} for the gradient, a combination of strategies was employed. Primarily, local observables were use to combat vanishing gradients and improve the trainability of the algorithm, as discussed in the paper ``Cost function dependent barren plateaus in shallow parameterized quantum circuits'' \cite{Cerezo_2021}.
Additionally, leveraging the concepts presented in the paper titled ``An initialization strategy for addressing barren plateaus in parameterized quantum circuits'' \cite{Grant_2019}, if the accuracy of an initial guess fell below a specific threshold, an alternative point in the parameter space was selected. By incorporating this valuable technique,  a favorable starting point within the optimization landscape could be selected and  training become feasible.

\subsection{Architecture}

In the QNN, each layer applies a transformation to the quantum state, generating a new state as input for the subsequent layer. The parameters of each layer are optimized through training. This approach based on the paper ``Data re-uploading for a universal quantum classifier''\cite{P_rez_Salinas_2020}, allows the quantum classifier to handle complex data with multiple input dimensions and output categories. 

The QNN class, implemented as a PyTorch module, allows for the specification of the number of layers and features during initialization. This design choice enables a modular framework that can be readily adjusted and modified during hyper-parameter tuning to align with the underlying geometry of the problem. For each architecture that was tested, a distinct class was implemented, and the corresponding code can be found in the GitHub repository \cite{githublink}.

Selecting the depth of the QNN involves a trade-off between computational resources and performance. Deeper QNNs may provide enhanced representation power, but they also introduce increased optimization complexity due to the requirement of either more qutrits or longer gate sequences. An increase in depth carries, also, the potential risk of over-fitting, necessitating a delicate balance to be struck in model design. 

\begin{figure}[ht]
\begin{mybox}
\begin{center}
\begin{quantikz}
\lstick{\ket{\textbf{0}}} & \gate{\textbf{U}(\vec{x})} & \phase{|\boldsymbol{\psi}_{\vec{x}}\rangle} & \gate{\textbf{U}(\vec{w})} & \phase{|\boldsymbol{\psi}_{\vec{x}, \vec{w} }\rangle} & \meter{}  \qw
\end{quantikz}
\end{center}
  \caption{ \label{fig:qnnlayer} A Single QNN Layer composed of Encoding and Variational Layers}
  \end{mybox}
\end{figure}

As shown in Fig.~\ref{fig:qnnlayer}, a single layer of the QNN comprises an encoding layer and a variational layer. The encoding layer, denoted as $\boldsymbol{U(\vec{x})}$, follows a feature map similar to the kernel.

\begin{figure}[ht]
\begin{mybox}
\begin{center}
\begin{quantikz}
\lstick{\ket{\boldsymbol{\psi}}} & \gate{\textbf{H}} &\gate{\textbf{R}_{\boldsymbol{\lambda_1}}(x_{1})}&\gate{\boldsymbol{\boldsymbol{R_{\lambda_2}}}(x_{2})}&\gate{\boldsymbol{R_{\lambda_3}}(x_{3})}&\gate{\boldsymbol{R_{\lambda_4}}(x_{4})}  & \ket{\boldsymbol{\psi_{\vec{x}}}}  \qw
\end{quantikz}
\end{center}
  \caption{\label{fig:encodlayer} The Encoding Layer $\textbf{U}(\vec{x})$ of a QNN layer made of a single qutrit}
  \end{mybox}
\end{figure}

As illustrated in Fig.~\ref{fig:encodlayer}, the initial step involves preparing the qutrit in a superposition state by applying the Hadamard operator. Subsequently, the features are encoded onto the qutrit using the first four Gell-Mann matrices. In the following variational layer, as depicted in Fig.~\ref{fig:varlayer}, the remaining four Gell-Mann matrices are utilized, and the weights of the QNN's parameters are optimized.

\begin{figure}[ht]
\begin{mybox}
\begin{center}
\begin{quantikz}
\lstick{\ket{\boldsymbol{\psi_{\vec{x}}}}} & 
\gate{\boldsymbol{R_{\lambda_5}}(w_{1})}&\gate{\boldsymbol{R_{\lambda_6}}(w_{2})}&\gate{\boldsymbol{R_{\lambda_7}}(w_{3})}&\gate{\boldsymbol{R_{\lambda_8}}(w_{4})} & \ket{\boldsymbol{\psi_{\vec{x}, \vec{w} }}} \qw
\end{quantikz}
\end{center}
  \caption{\label{fig:varlayer} The Variational Layer $\textbf{U}(\vec{w})$ of a QNN layer made of a single qutrit}
  \end{mybox}
\end{figure}

Fig.~\ref{fig:qnnlayer} represents a single layer of the QNN for a single qutrit. Additional layers are stacked at the end, similar to how perceptrons are added in classical NNs, to increase its depth. The effectiveness of increasing the number of qutrits varied across datasets. In some cases, utilizing two qutrits led to improved performance, while in others, a single qutrit was sufficient. We note here that using more than two qutrits was unnecessary for the datasets under testing.

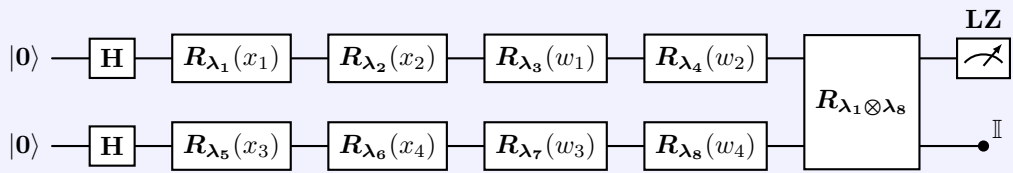
\begin{figure}[ht]
\begin{mybox}
\begin{center}
\begin{quantikz}
\lstick{\ket{\textbf{0}}} &\gate{\textbf{H}}&\gate{\boldsymbol{R_{\lambda_1}}(x_{1})}&\gate{\boldsymbol{R_{\lambda_2}}(x_{2})}&\gate{\boldsymbol{R_{\lambda_3}}(w_{1})}&\gate{\boldsymbol{R_{\lambda_4}}(w_{2})}&\gate[2]{\boldsymbol{R_{\lambda_1\otimes\lambda_8}}}& \meter{\textbf{LZ}} \\
\lstick{\ket{\textbf{0}}} &\gate{\textbf{H}}&\gate{\boldsymbol{R_{\lambda_5}}(x_{3})}&\gate{\boldsymbol{R_{\lambda_6}}(x_{4})}&\gate{\boldsymbol{R_{\lambda_7}}(w_{3})}&\gate{\boldsymbol{R_{\lambda_8}}(w_{4})}&& \phase{\mathbb{I}}
\end{quantikz}
\end{center}
  \caption{ \label{fig:qnn2qtrt} A single layer of the QNN with two qutrits using all eight Gell-Mann rotations and an entangling operator.}
  \end{mybox}
\end{figure}

The architecture presented in Fig.~\ref{fig:qnn2qtrt}, resembling the feature map for two qutrits in the quantum kernel section, employs the Gell-Mann matrices in a similar manner. However, a  difference arises in the type of the entangling operator that we employ:
\begin{equation}
\boldsymbol{R_{\lambda_1\otimes\lambda_8}} = \exp[i\boldsymbol{\lambda_1}\otimes\boldsymbol{\lambda_8}].
\end{equation}
 While this entangling operation exhibited superior performance compared to other types explore, its impact to the task's success ratio was depended on the underlying structure of the dataset, emphasizing its data-specific nature.

Finally, for the QNN a local measurement approach is used, employing the $\boldsymbol{LZ}$ operator. 
For the sake of simplicity in the training process, the probabilities of the quantum state are utilized instead of running the circuit multiple times and obtaining the expectation values. This is made possible by accessing the quantum state directly through the Python simulation. For each architecture that was tested, a distinct class was implemented, and the corresponding code can be found in the GitHub repository \cite{github}.

\subsection{Results}

During the training process, a thorough search was conducted to explore various combinations of learning rates, weight decay, and momentum factors in the optimizers that supported this. The objective was to fine-tune the model's hyper-parameters and discover the optimal configuration that maximized convergence speed.

In the preliminary experiments, the parameter shift rule was employed, but it was later replaced with the \texttt{RMSProp} optimizer, which demonstrated superior performance. \texttt{RMSProp} was used to update the parameters of the QNN, with an initial \texttt{learning rate} set to 0.001, a decay factor \texttt{gamma} of 0.9, and an \texttt{epsilon} value of 1e-8 for numerical stability.

The input features were appropriately scaled, using scikit-learn's \texttt{StandardScaler} pre-processing class, which transforms the data by removing the mean and scaling to unit variance. The standard score of a sample $x$ is calculated as: $z = \frac{x - u}{s}$, where $u$ is the mean of the training sample, and $s$ is the standard deviation of the training samples. Standardization is a necessary requirement for neural networks, as they might behave poorly if the individual features do not resemble standard normally distributed data.

The \texttt{Dataset} and \texttt{DataLoader} classes were utilized, enabling efficient preprocessing, seamless integration of data loading with the model training process, and leveraging the capabilities of the broader PyTorch ecosystem. Various batch sizes were tested to strike the optimal balance between training efficiency and model performance. Due to the small size of the datasets, a small \texttt{batch size} of 4 was chosen, since increasing the batch size beyond this point proved to be impractical.

PyTorch's \texttt{cross entropy} loss function was utilized to compute the loss between the input logits and target, as it operates on the probabilities of the quantum state. The computed loss is then used to perform the backward pass, and update the model's parameters. 

\subsubsection{Binary Classification}

\begin{figure} [ht]
\begin{center}
\begin{tabular}{c} 
\includegraphics[height=5cm]{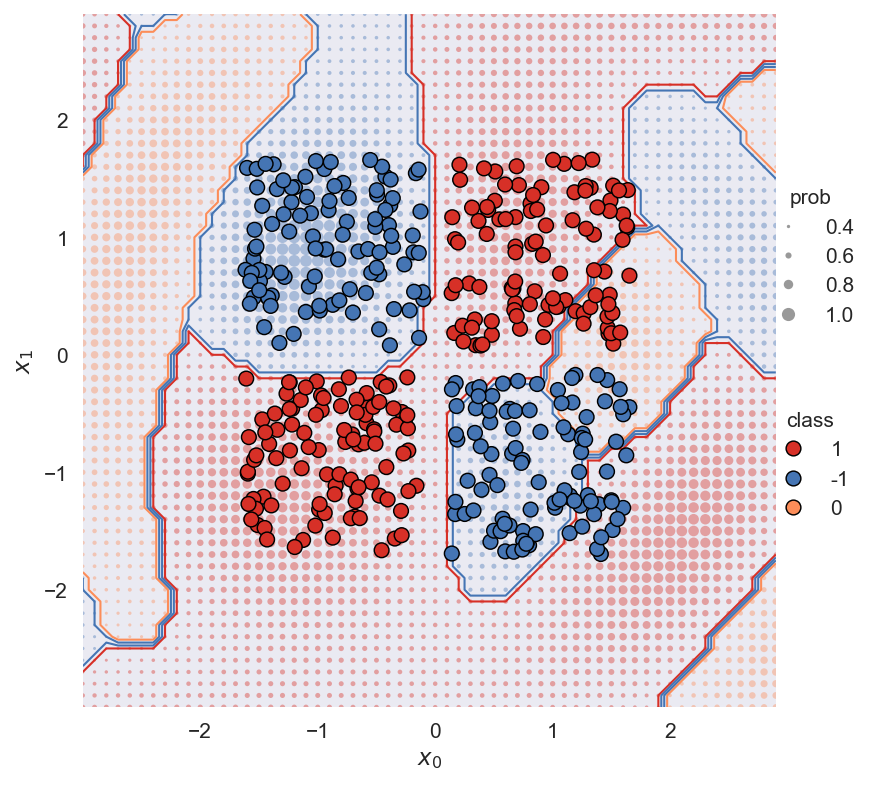}
\includegraphics[height=5cm]{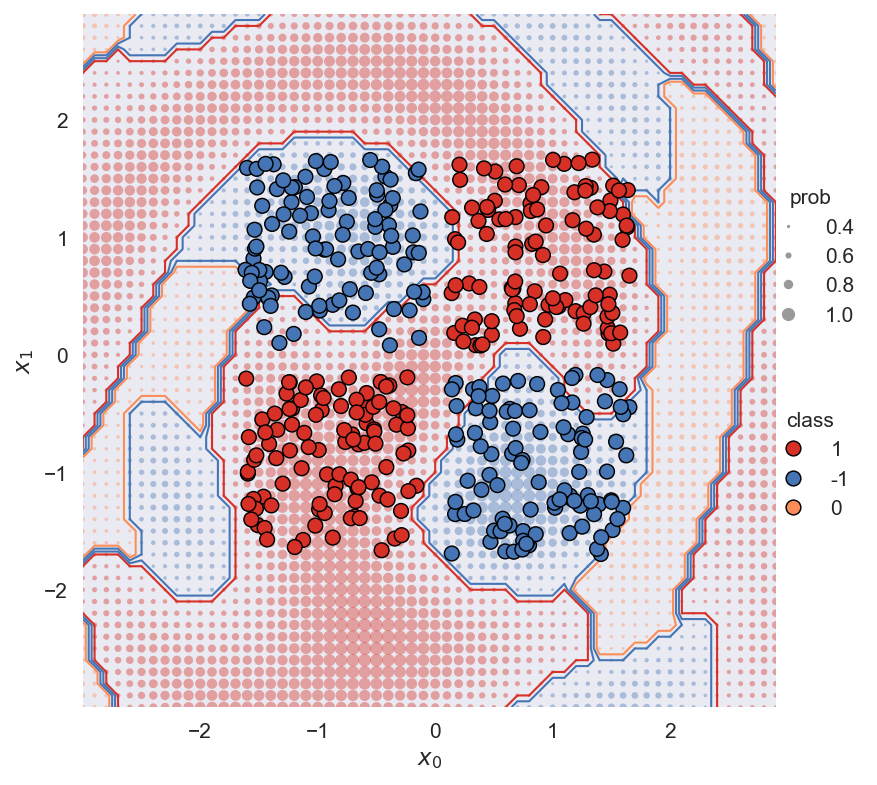}
\includegraphics[height=5cm]{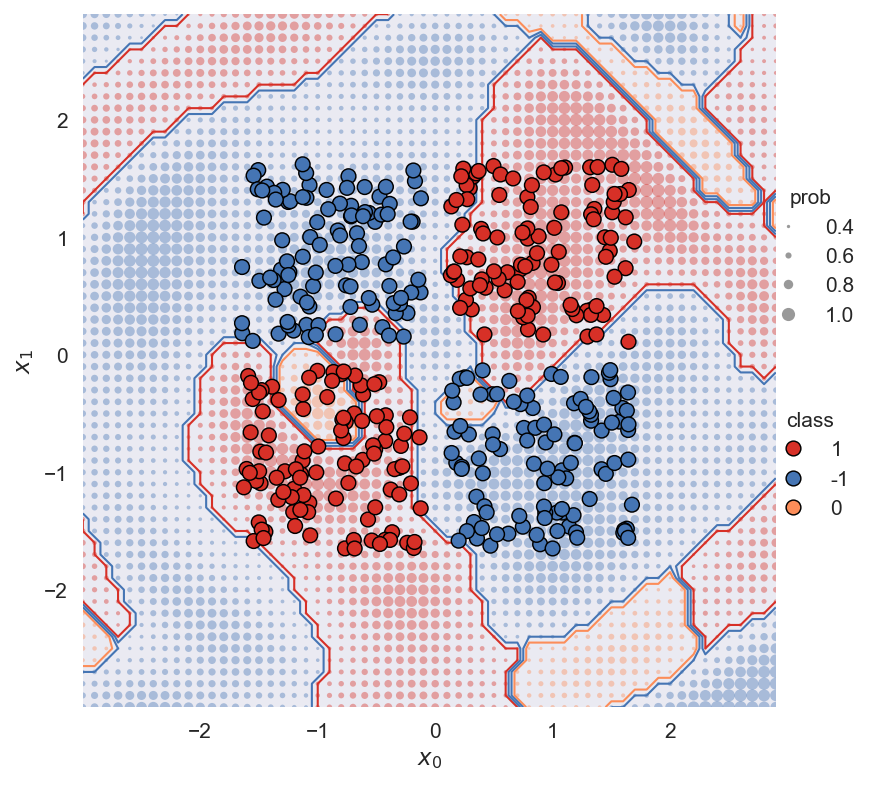}
\end{tabular}
\end{center}
\caption{ \label{fig:qnnxor} Decision Boundaries of QNN with a single qutrit on the XOR dataset. 1 Layer Accuracy: $85.3$. 2 Layers Accuracy: $91.25\%$. 3 Layers Accuracy: $100\%$ }
\end{figure} 

In the initial evaluation of the QNN's performance and to compare with the QKSVM's separation capacity, the XOR and Moons datasets were employed. Notably, there exists a distinction between NNs and support vector machines in terms of their approach to learning decision boundaries.
When considering artificial NNs or perceptrons, the separation capacity is determined by the number of hidden layers present in the network. Specifically, the absence of hidden layers restricts the network to learning only linear problems. However, the inclusion of one hidden layer enables the network to learn any continuous function, thus facilitating the adoption of arbitrary decision boundaries. 

This behavior was also detected in the case of the QNN, where an increase in depth resulted in the emergence of increasingly complex and intricate decision boundaries. This can be clearly observed in Fig.~\ref{fig:qnnxor}, which demonstrates the progression from 1 to 3 layers. Additionally, as the QNN learns to effectively separate the classes in the XOR dataset, the accuracy naturally improves, progressing from 85.3\% to a perfect 100\%.

\begin{figure} [ht]
\begin{center}
\begin{tabular}{c} 
\includegraphics[height=3.64cm]{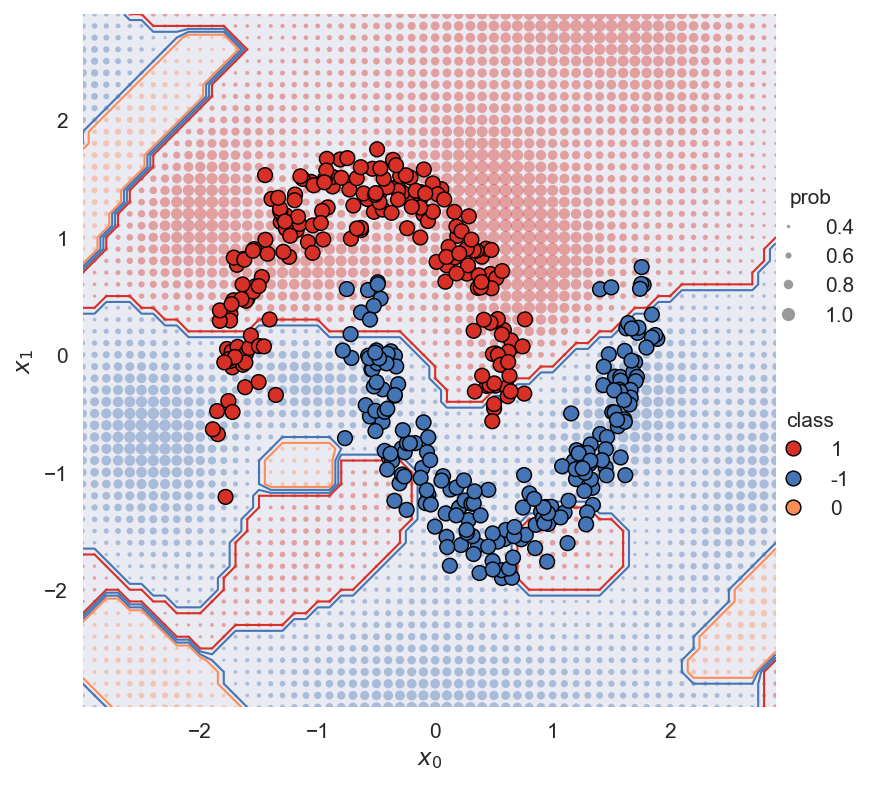}
\includegraphics[height=3.64cm]{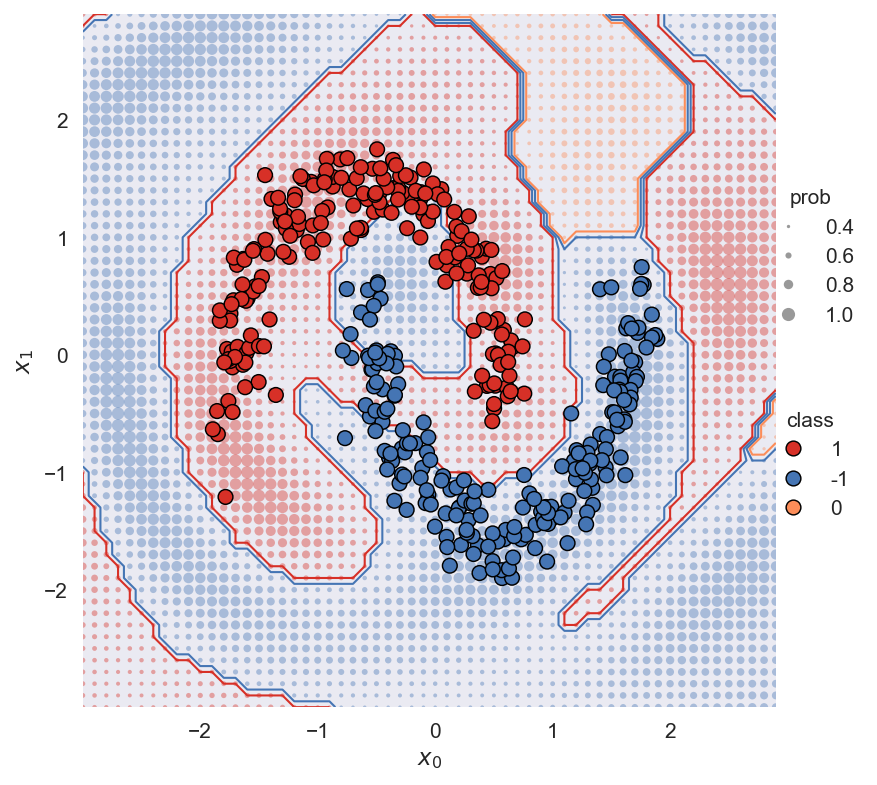}
\includegraphics[height=3.64cm]{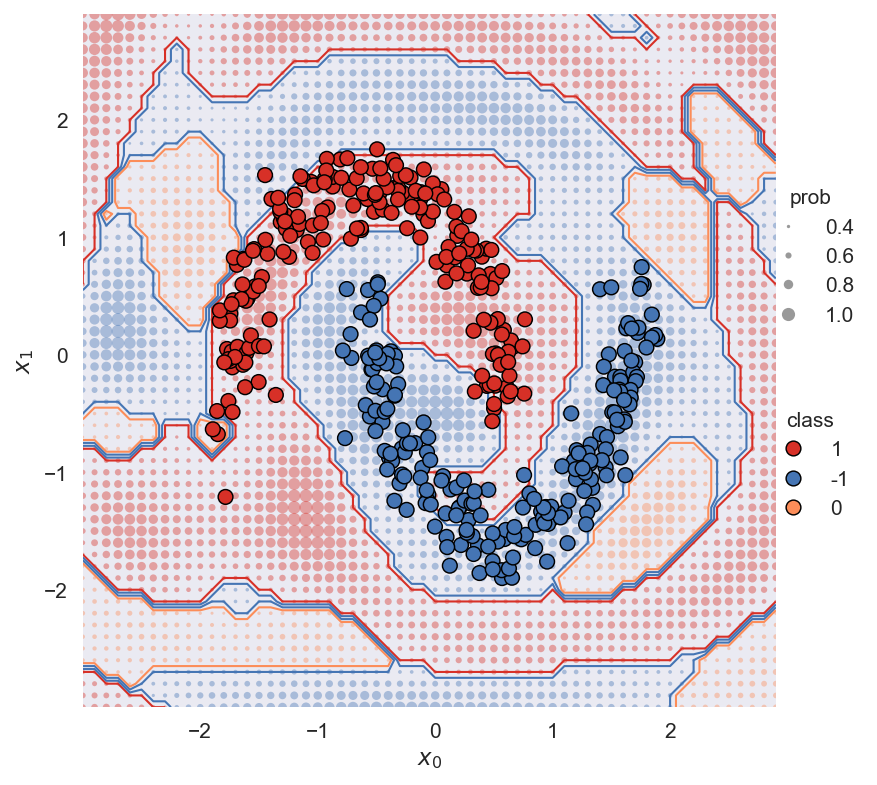}
\includegraphics[height=3.64cm]{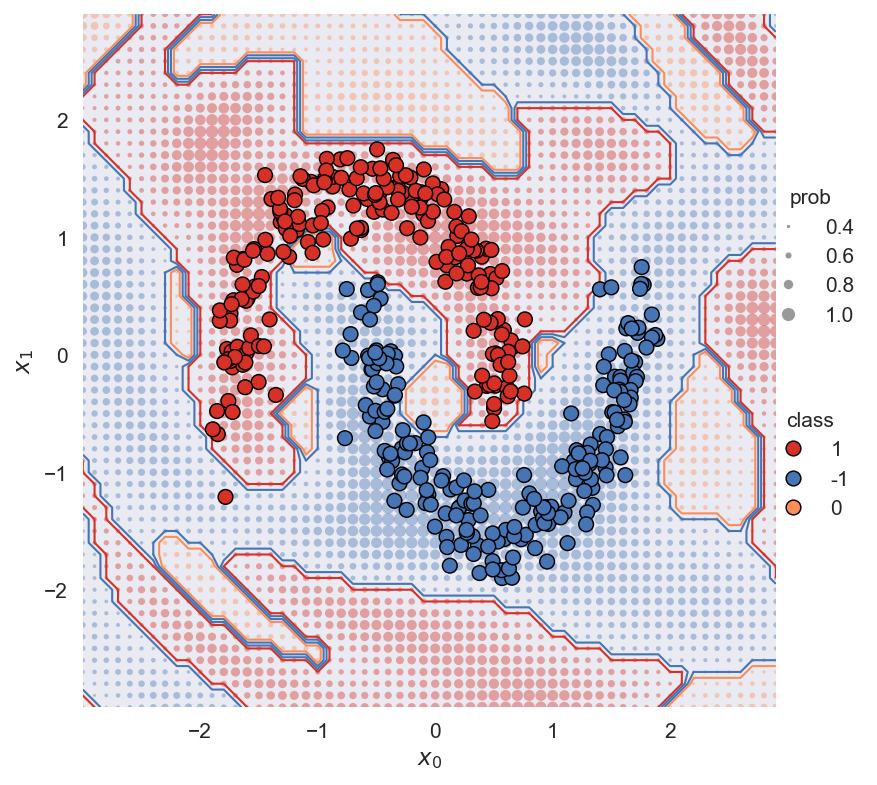}
\end{tabular}
\end{center}
\caption{ \label{fig:qnnmoons} Decision Boundaries of QNN with a single qutrit on the Moons dataset. 1 Layer Accuracy: $85.3$. 2 Layers Accuracy: $91.25\%$. 3 Layers Accuracy: $92.5\%$. 4 Layers Accuracy: $100\%$ }
\end{figure} 

The same phenomenon is even more evident in Fig.~\ref{fig:qnnmoons}, where the presence of one layer results in a simple linear separation, followed by increasing intricacy as additional layers are introduced. In this case, the decision boundaries progressively improve to better align with the interleaving shapes present in the Moons dataset.

The accuracy of the QNN on the Moons dataset exhibits a steady improvement, starting from 85.3\% with one layer and reaching 100\% with four layers. Similar to the quantum kernel's measurement in the previous section, the accuracy presented below the plots is determined using the test set, while the decision boundaries are plotted over the entire dataset, providing a comprehensive visualization.

For both the XOR and Moons datasets, the QNN employed a single qutrit architecture. These binary classification problems demonstrated that using more than one qutrit was unnecessary for effective learning. However, since each layer requires a minimum of four dimensions (see Fig.~\ref{fig:encodlayer}), the two-dimensional features were ``re-uploaded'' in the subsequent Gell-Mann rotations.

\subsubsection{Multiclass Classification}

\begin{figure} [ht]
\begin{center}
\begin{tabular}{c} 
\includegraphics[height=4cm]{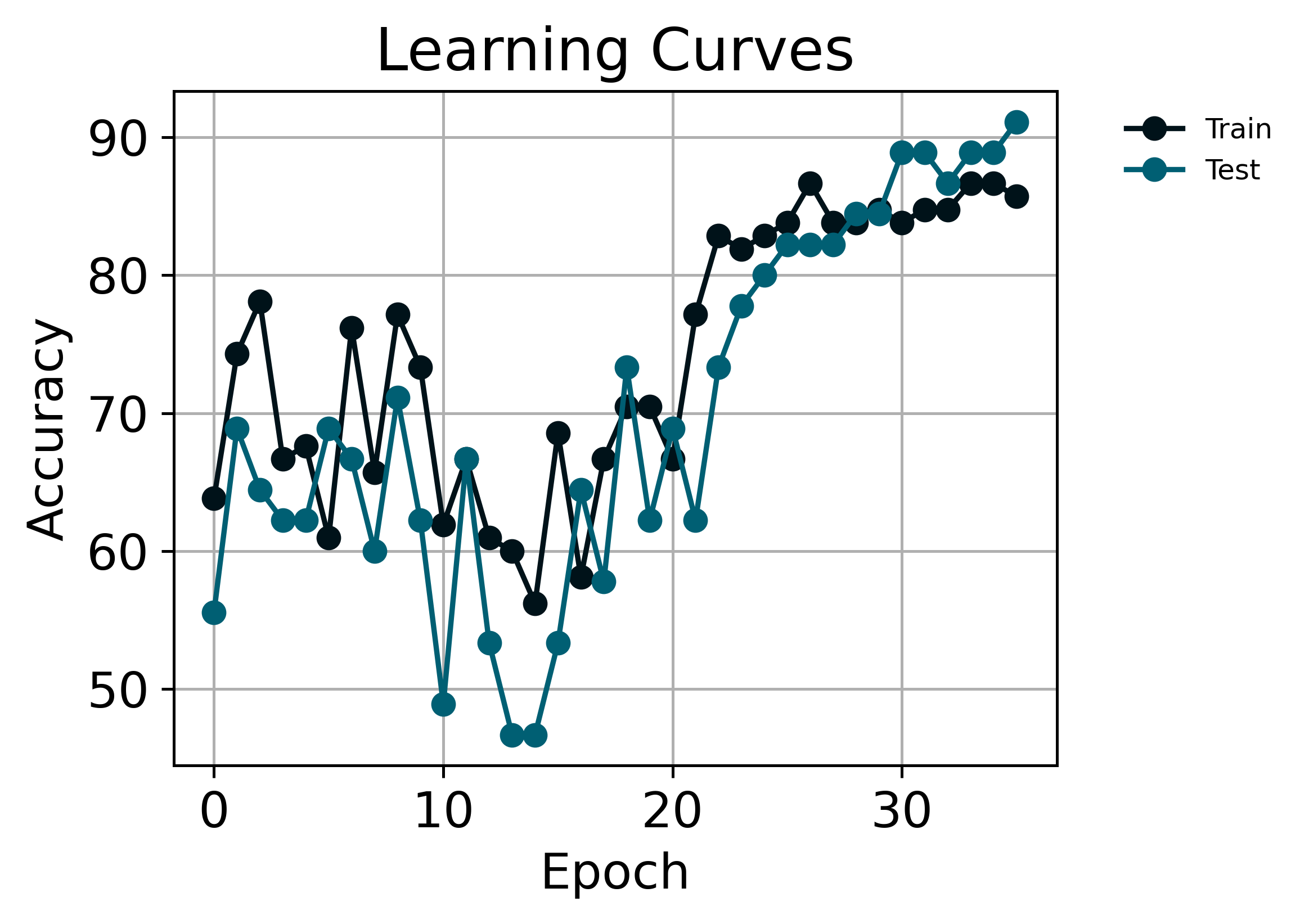}
\includegraphics[height=4cm]{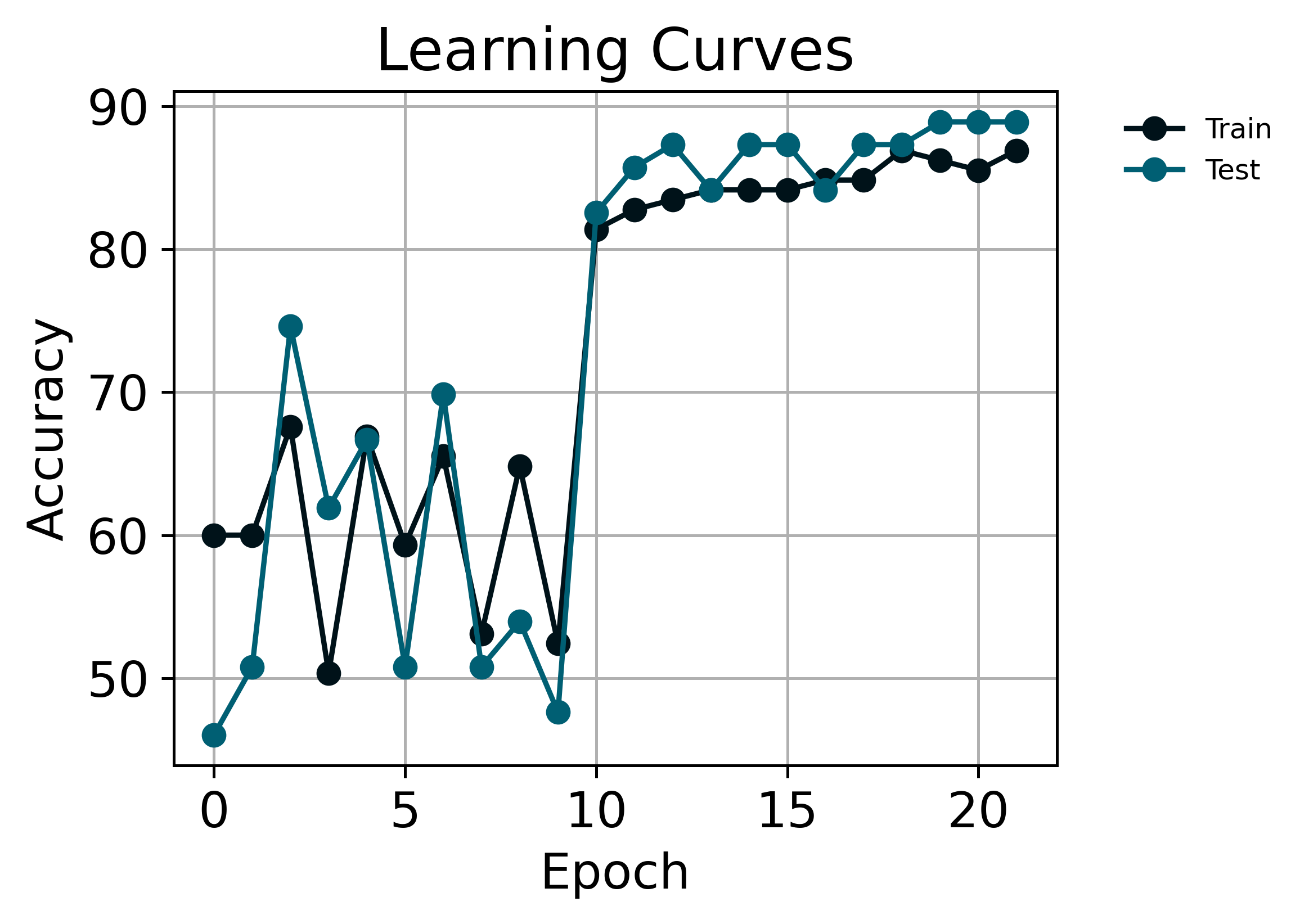}
\includegraphics[height=4cm]{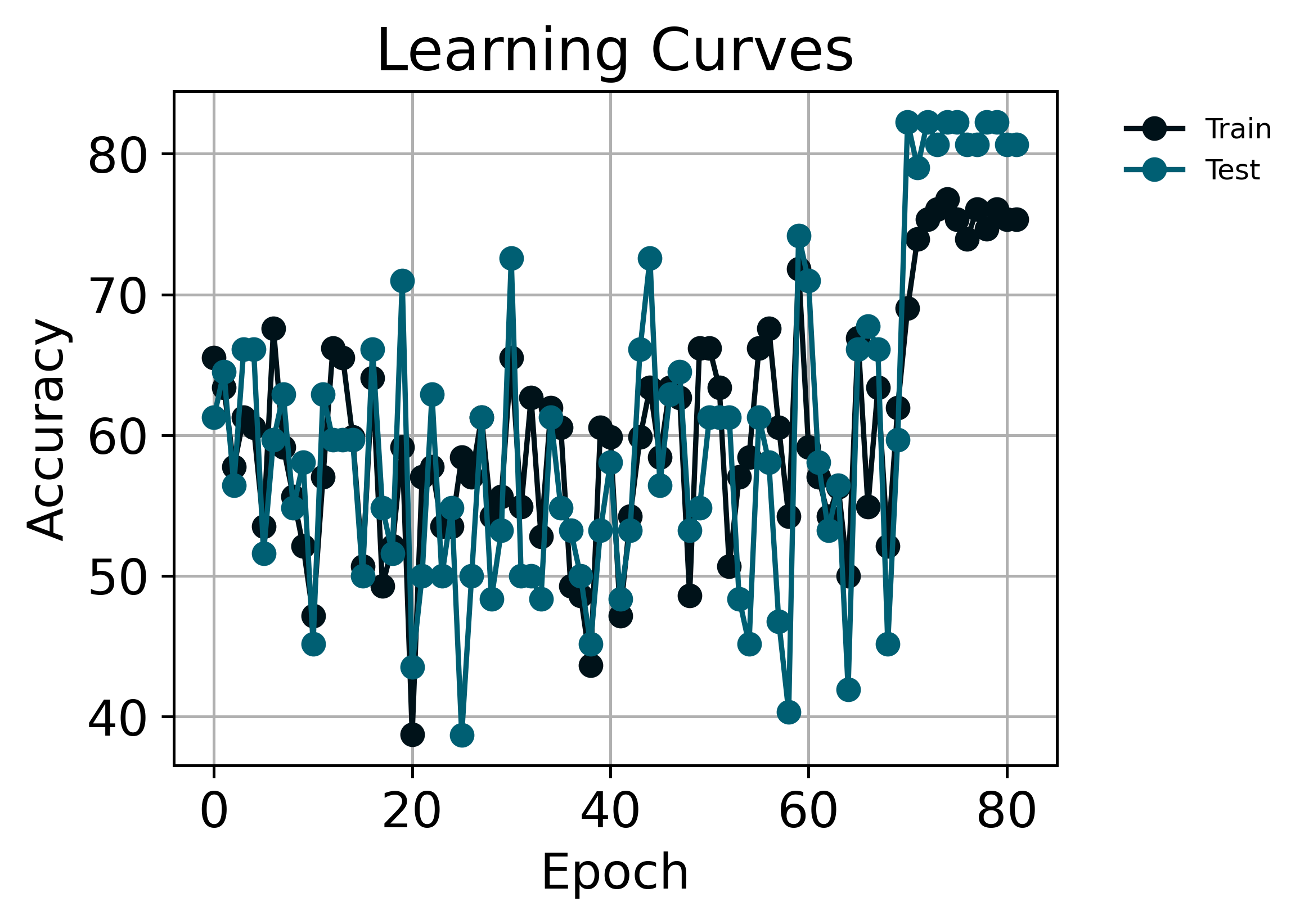}
\end{tabular}
\end{center}
\caption{ \label{fig:lc} Learning Curves for single qutrit QNN with 4 layers for IRIS dataset, 3 layers for SEED dataset and 2 layers for GLASS.}
\end{figure}

For the multiclass classification tasks, learning curves are plotted Fig.~\ref{fig:lc} to visually represent the model's performance and learning progress over time. These curves reveal important trends such as convergence and potential over-fitting or under-fitting.

The Iris dataset consists of four features, while the remaining datasets have higher dimensions. To enhance computational efficiency while maintaining consistent encoding, PCA was employed to reduce the dimensions to four in all datasets, except for the Glass dataset. In the Glass dataset, the original dimensions were retained and encoded in the subsequent layer, since this yielded better accuracy.

The  results in this section are obtained using the single qutrit QNN architecture. Although the quantum kernel achieved better accuracy with two qutrits compared to the single qutrit in multiclass problems, the QNN achieved similar performance with both architectures. Considering the longer training time required by the two qutrit architecture, which is expected due to the increased number of operations, the decision was made to utilize the single qutrit architecture for all tasks.

In the Iris dataset, the model achieves its optimal performance with four layers, as demonstrated by the learning curves presented in Fig.~\ref{fig:lc}, having a 91.11\% accuracy on the test set. The increased depth of the QNN allowed the model to effectively capture the intricate features and complex patterns within the data. 

For the Seed dataset, the QNN requires fewer epochs to train effectively. As illustrated in Fig.~\ref{fig:lc}, the model achieves its peak performance with three layers, attaining an accuracy of 88.89\%, a macro average F1-Score of 88.96\%, and similar performance across Recall and Precision metrics. 

As mentioned earlier, the Glass dataset utilized all available features, necessitating an increase in the number of layers and parameters for the variational layer. Consequently, more epochs are required to achieve a good initial starting point and optimize the model. This prolonged convergence is particularly noticeable in the case of the two-layer architecture, where it took 70 epochs to find a favorable state.

Lastly, for the Wine dataset, the model necessitates an extended number of epochs to identify an optimal initial starting point. Interestingly, the model demonstrates comparable accuracy with both two and three layers. Specifically, the accuracy achieved with two layers is recorded at 85.19\%, and exhibiting similar performance across all metrics.

The observed initial instability in the learning curves can be attributed to the initialization method, whereby the model randomly selects a point in the hyper-parameter space. While this approach has demonstrated usefulness, it is worth considering alternative strategies such as employing a quantum natural gradient or leveraging Fisher information \cite{Liu_2019}. These methods have the potential to enhance the training landscape, mitigating the need for ad hoc measures, and provide more robust gradient techniques.

\begin{table}[ht]
\caption{Metrics of the single qutrit QNN's performance on multiclass datasets.}
    \label{tab:accuracyqnn}
    \begin{center}
    \begin{tabular}{|l|l|l|l|l|}
    \hline
     & IRIS    & WINE    & GLASS   & SEED    \\ \hline
    Recall             & 91.11\% & 84.23\% & 82.54\% & 88.89\% \\ \hline
    Precision          & 91.23\% & 84.61\% & 82.96\% & 89.10\% \\ \hline
    Accuracy           & 91.11\% &  85.19\% & 82.54\% & 88.89\% \\ \hline
    F1-score           & 91.10\% & 84.37\% & 82.35\% & 88.96\% \\ \hline
    Layers           & 4 & 3 & 2 & 2 \\ \hline
    \end{tabular}
     \end{center}
\end{table}

Table ~\ref{tab:accuracyqnn}, provides an overview of the QNN's performance, presenting the comprehensive metrics obtained from the aforementioned benchmarks, with the respective number of layers for each dataset. Multiple benchmarks were conducted with comparable outcomes, and models that exhibited superior performance, but suffered from over-fitting were excluded. The table thus includes the best-performing models that strike a balance between performance and generalization. 

\section{Comparative Analysis}

\subsection{QKSVM vs. QNN}

The visual representation of decision boundaries for two-dimensional data, see Figs.~\ref{fig:qksvm1}-\ref{fig:qksvm2},\ref{fig:qnnxor}-\ref{fig:qnnmoons}, proved to be insightful in understanding the geometric structure of the solutions offered by different models. It shed light on the models' capacity, highlighting notable distinctions between the fitting processes of QKSVM and QNN.

\begin{figure} [ht]
\begin{center}
\begin{tabular}{c} 
\includegraphics[height=5cm]{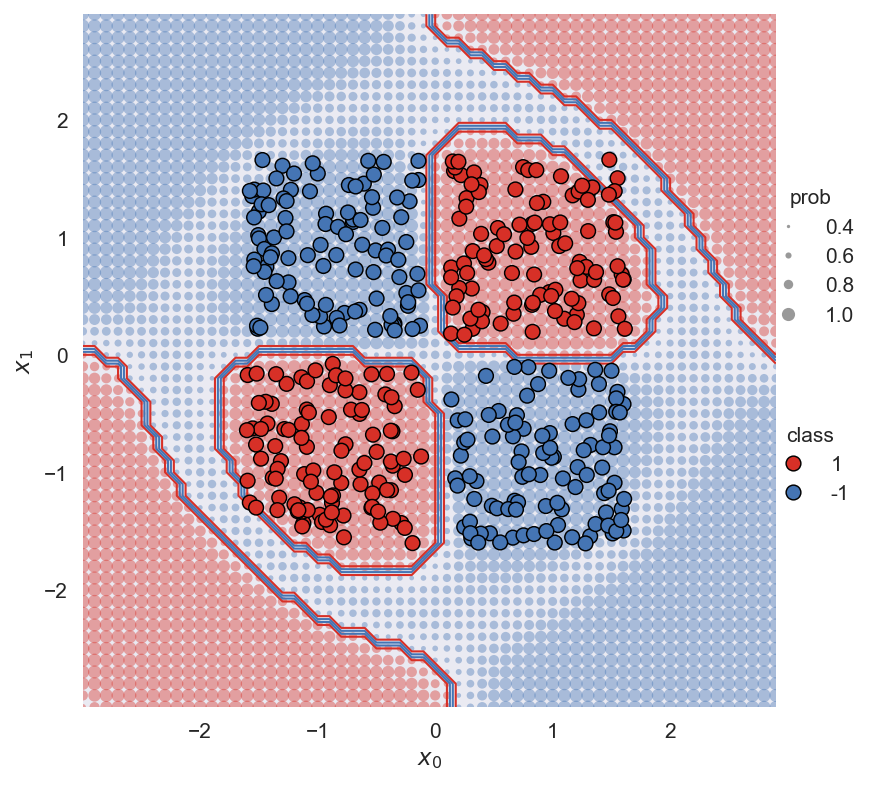}
\includegraphics[height=5cm]{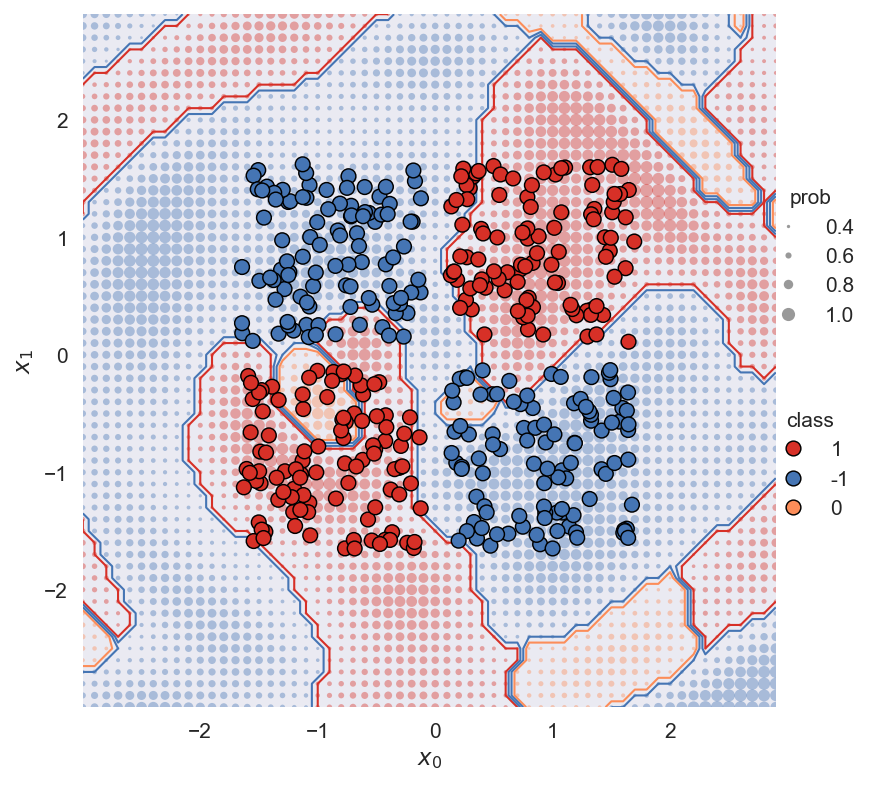}
\end{tabular}
\end{center}
\caption{ \label{fig:qnnxor2} XOR Decision Boundaries Comparison of QKSVM of two qutrits and single qutrit QNN with 3 layers.}
\end{figure} 

\begin{figure} [ht]
\begin{center}
\begin{tabular}{c} 
\includegraphics[height=5cm]{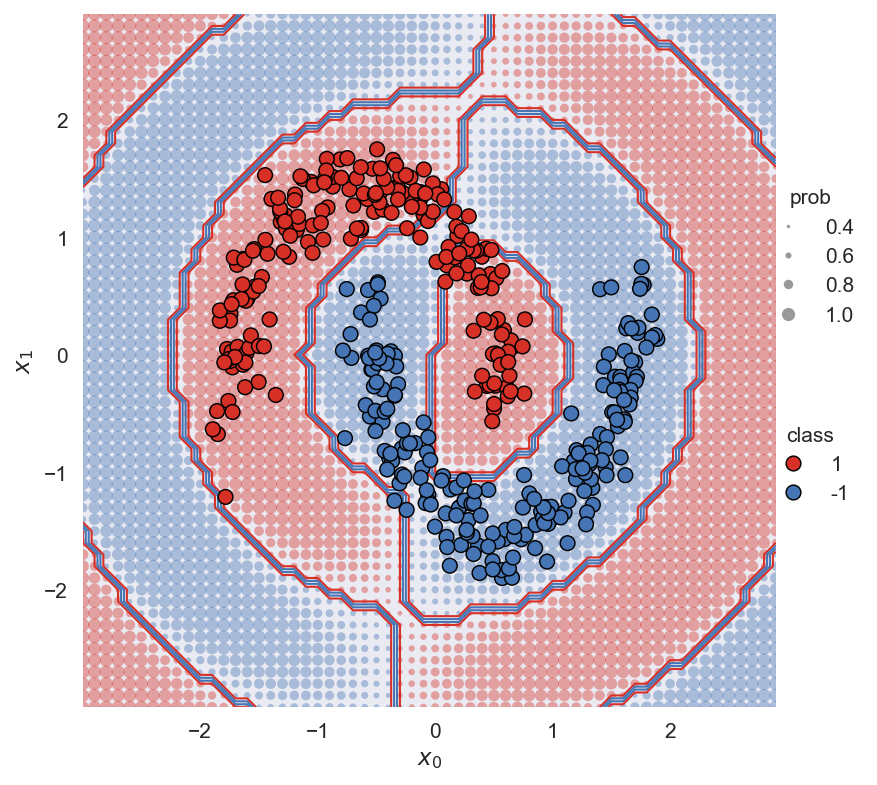}
\includegraphics[height=5cm]{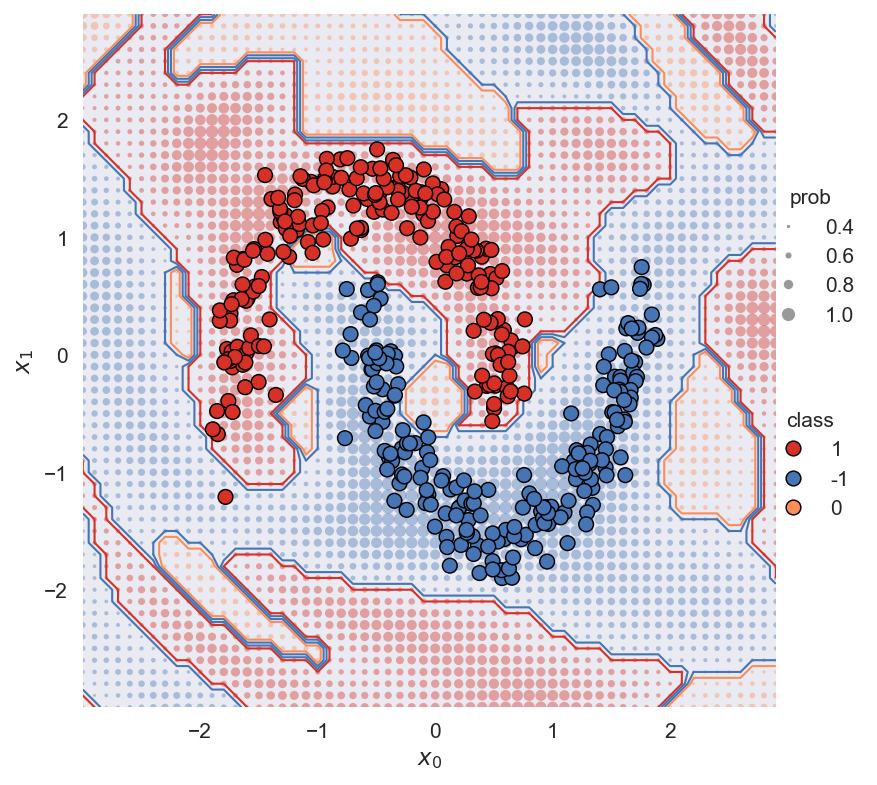}
\end{tabular}
\end{center}
\caption{ \label{fig:qnnmoons2} Moons Decision Boundaries Comparison of QKSVM of two qutrits and single qutrit QNN with 4 layers.}
\end{figure} 

With the plots Figs.~\ref{fig:qnnxor2}-\ref{fig:qnnmoons2} we put side-by-side the geometric structure of the solutions achieved by QKSVM and QNN. One may observe that the decision boundaries derived with QKSVM exhibit a distinct and easily interpretable geometric separation in both the Moons and XOR datasets. In contrast, the decision boundaries produced by the QNN are characterized by intricate and nuanced patterns that appear to closely align with the data points, particularly evident in the case of the Moons dataset. This behavior aligns with the inherent capability of NNs to capture intricate and non-linear relationships, enabling them to model more effectively complex patterns within the datasets.
This distinguishing characteristic positions QNNs as promising tools for addressing real-world problems where datasets lack explicit geometric structures. Their potential lies in their ability to capture and exploit the inherent complexity and non-linearity embedded within the data.

\begin{table}[ht]
    \caption{Comparison of Metrics for QKSMV and QNN on Multiclass Datasets}
    \label{tab:comparison}
    \begin{center}
    \begin{tabular}{|l|l|l|l|l|}
    \hline
    & \textbf{IRIS} & \textbf{WINE} & \textbf{GLASS} & \textbf{SEED} \\
    \hline
    \textbf{Recall (QKSMV)} & 90.00\% & 83.97\% & 90.48\% & 88.10\% \\
    \hline
    \textbf{Precision (QKSMV)} & 92.31\% & 83.51\% & 90.43\% & 89.56\% \\
    \hline
    \textbf{Accuracy (QKSMV)} & 90.00\% & 83.33\% & 90.24\% & 88.10\% \\
    \hline
    \textbf{F1-score (QKSMV)} & 89.77\% & 83.42\% & 90.21\% & 87.93\% \\
    \hline
    \textbf{Recall (QNN)} & 91.11\% & 84.23\% & 82.54\% & 88.89\% \\
    \hline
    \textbf{Precision (QNN)} & 91.23\% & 84.61\% & 82.96\% & 89.10\% \\
    \hline
    \textbf{Accuracy (QNN)} & 91.11\% & 85.19\% & 82.54\% & 88.89\% \\
    \hline
    \textbf{F1-score (QNN)} & 91.10\% & 84.37\% & 82.35\% & 88.96\% \\
    \hline
    \end{tabular}
    \end{center}
\end{table}

Table ~\ref{tab:comparison} with the multiclass results provides further evidence of the QNN's prowess in uncovering and leveraging intricate patterns within real-world datasets. In all evaluated datasets, the QNN consistently outperformed the quantum kernel across all metrics, except for the Glass dataset. This outcome underscores the QNN's superior performance and its ability to effectively model complex relationships and patterns present in the data.

It is important to note that, in the case of QKSVM, utilizing the original features of the dataset led to superior outcomes compared to applying PCA and using the reduced features as input. Interestingly, the opposite trend was observed in the context of the QNN.
This difference in performance can be attributed to several factors, which require further investigation and analysis. The QKSVM might be more effective when the data distribution aligns well with the quantum feature space. Naturally, it might be the case that the increased depth of the QNN and number of parameters simply required a different optimization approach.    
The size of the dataset can also play a role in the observed results. If the dataset is relatively small, which is the case with the datasets which we have used, QKSVM might be more resilient to over-fitting and better able to capture the underlying patterns in the raw features, whereas the QNN could struggle due to the increased model complexity.

\subsection{Single qutrit QNN vs. VQC with four qubits vs. Classical SVM}

In order to evaluate the performance of the qutrit QNN, a comparative study is conducted using a qubit VQC and a classical SVM. As a reference we also include in Table~\ref{tab:comparisonclassic} the classification results of the QKSVM method using a two-qutrit feature map.

Scikit-learn's SVC is selected as a representative classical ML method due to its versatility and suitability as a baseline classifier for model evaluation. By conducting a comparative analysis of the performance between quantum algorithms and scikit-learn's SVC, the strengths and weaknesses of quantum approaches in tackling multiclass classification problems can be assessed. 

To implement the qubit VQC, Qiskit's VQC is utilized, which constructs a variational (parametrized) quantum circuit for data classification. The VQC in Qiskit offers support for various loss functions and optimizers, allowing flexibility in the training process. It also provides the option for warm start, which means that weights from a previous fit can be used to initialize the next fit. 
Qiskit's VQC takes as parameters the number of qubits, feature map, ansatz, loss function, and optimizer. For this study, we used four qubits, matching the number of features in the Iris dataset. In the other datasets, PCA was employed to reduce the number of features to four. From optimizers the L\_BFGS\_B\cite{lbfgs} and COBYLA\cite{cobyla} were selected. As the feature map,
the ZZFeatureMap was selected, a commonly used data encoding method. For the ansatz, the RealAmplitudes circuit was employed, which consists of alternating layers of rotations and CNOT gates.
Both the feature map and the variational quantum circuit can be repeated a certain number of times, which is adjustable and serves as an argument in the circuit's construction. This provides flexibility in designing the architecture of the circuit.
    
\begin{table}[ht]
\caption{Comparison of the qutrit QNN's performance to Qubit VQC and SVM}
    \label{tab:comparisonclassic}
    \begin{center}
    \begin{tabular}{|l|l|l|l|l|}
    \hline
    & \textbf{IRIS} & \textbf{WINE} & \textbf{GLASS} & \textbf{SEED} \\
    \hline
    \textbf{Classical SVM Accuracy} & 100\% & 91.7\% & 100\% & 90.48\% \\
    \hline
    \textbf{QKSVM Accuracy} & 90.00\% & 83.33\% & 90.24\% & 88.10\% \\
    \hline
    \textbf{Qutrit QNN Accuracy} & 91.11\% & 85.19\% & 82.54\% & 88.89\% \\
    \hline
    \textbf{Qubit VQC Accuracy} & 86.67\% & 80.78\% & 68.98\% & 73.81\% \\
    \hline
    \textbf{Qutrit QNN Layers} & 4 & 2 & 2 & 3 \\
    \hline
    \textbf{Qutrit QNN Parameters} & 16 & 8 & 8 & 12 \\
    \hline
    \textbf{Qubit VQC ZZ Repetitions} & 1 & 1 & 2 & 2 \\
    \hline
    \textbf{Qubit VQC Parameters} & 12 & 16 & 20 & 24 \\
    \hline
    \end{tabular}
    \end{center}
\end{table}

The classical SVM achieves perfect accuracy on the IRIS and GLASS datasets, while for the WINE and SEED datasets, 91.7\% and 90.48\% respectively. This showcases the strong performance in multiclass classification tasks. The qutrit QNN achieves a competitive accuracy, demonstrating its potential in multiclass classification tasks, but it is still lower comparable to the classical SVM. It is worth noting that quantum methods, although promising are still ``green'' and are expected to have lower outright accuracy compared to classical methods.

In comparison, the QKSVM with two qutrits following the architecture of Fig.~\ref{fig:2qutfmap}, achieves slightly lower accuracy across all datasets and the qubit VQC's performance falls behind all of them. The circuit employing four qubits achieves 86.67\% accuracy on the IRIS dataset, 80.78\% on the WINE dataset, 68.98\% on the GLASS dataset, and 73.81\% on the SEED dataset. This suggests the need for further improvement, such as increasing the number of qubits or parameters, in order to enhance the performance.

In terms of model complexity, the qutrit QNN has 16 parameters for the IRIS dataset, 8 parameters for the WINE and GLASS datasets, and 12 parameters for the SEED dataset. On the other hand, the qubit VQC uses 1 repetition for the ZZ feature map on the IRIS and WINE datasets, and 2 repetitions on the GLASS and SEED datasets. The number of parameters in the ansatz of the VQC is 12 for the IRIS dataset, 16 for the WINE dataset, 20 for the GLASS dataset, and 24 for the SEED dataset.

These results highlight that even with fewer parameters, the single qutrit circuit outperforms the qubit VQC and can capture more information through its rotations. Additionally, the qubit circuit employs 4 qubits compared to the 1 qutrit of the QNN. Although increasing the number of qubits or parameters would likely improve the VQC to achieve comparable results, it would also increase the cost. The objective here was to demonstrate the performance of the VQC with a similar number of parameters. 

\begin{figure}[ht]
  \begin{center}
  \includegraphics[width=6.2in]{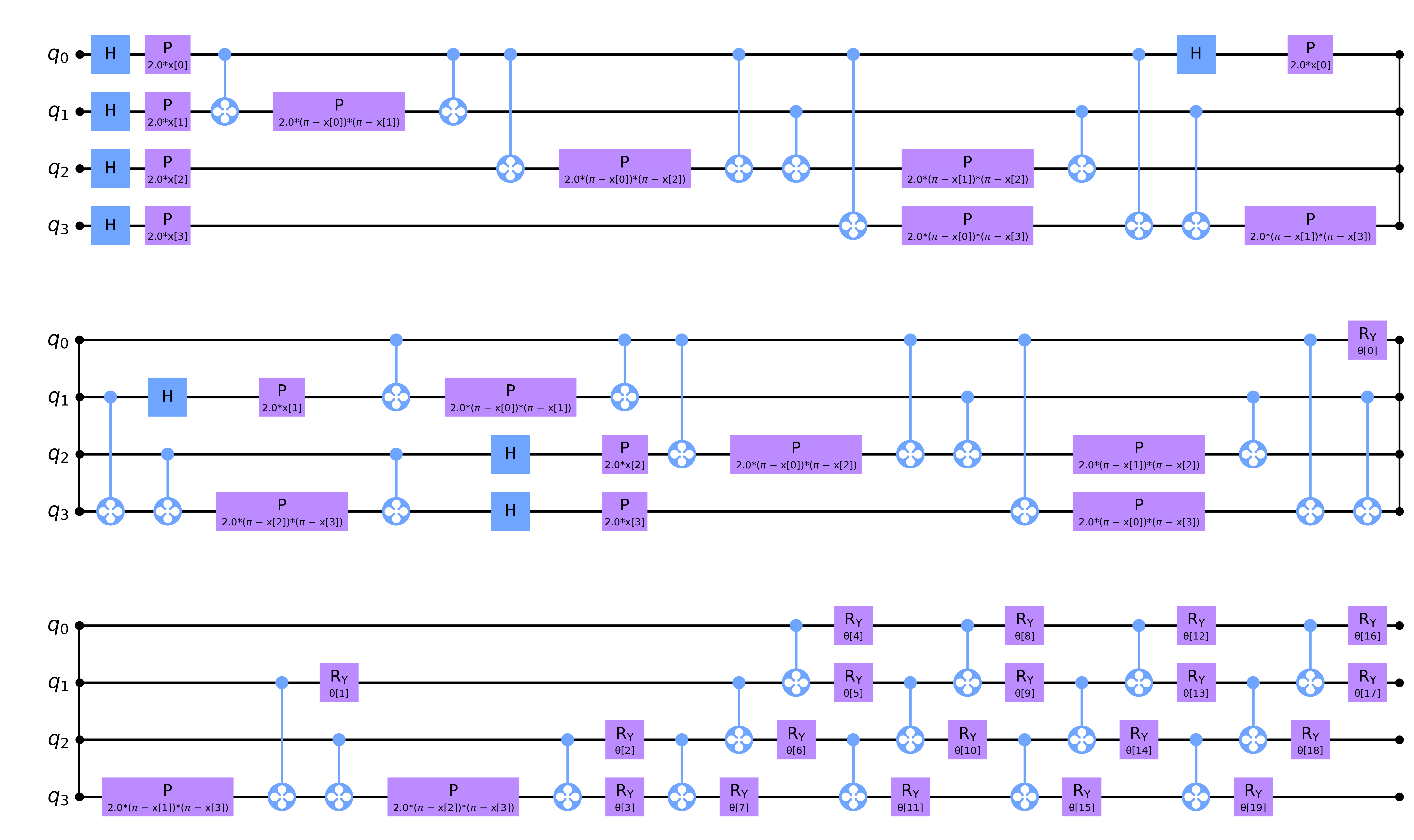}
  \caption{\label{fig:vqcqubit} Qubit VQC  utilizing the ZZ Feature Map ($2$ repetitions) and Real Amplitudes Ansatz on the variational circuit ($4$ repetitions). Each input data is loaded in the quantum circuit $8$ times.}
  \end{center}
\end{figure}

An overview of the complexity and number of gates required for the qubit VQC is given with Fig.~\ref{fig:vqcqubit}, specifically in the case of the Glass dataset which appears to be the most challenging task. This circuit yielded the highest accuracy for the dataset by utilizing 2 repetitions of the ZZ feature map and 4 repetitions of the variational circuit, resulting in a total of 20 trainable parameters. This circuit also took longer to train, and under-performed as compared to qutrit QNN, -even when varying the optimization algorithms.

The key point remains that despite having fewer parameters, the expanded computational space of the qutrit and the utilization of the Gell-Mann feature map are capable of capturing information that surpasses the capabilities of a circuit with four qubits. Potential improvements to the efficiency of  the qubit VQC involve increasing the repetitions of the ZZ feature map and the number of rotations or entangling operations. However, such modifications would introduce an unfair comparison to the shallow depth of the single qutrit QNN, which could also achieve higher accuracy by increasing its depth. 

These insights significantly contribute to advancing our understanding of the capabilities and limitations of quantum methods in the context of multiclass classification problems. Among the evaluated quantum approaches, the single qutrit QNN demonstrated competitive accuracy, indicating its potential effectiveness in tackling complex classification tasks. The QKSVM approach showcased comparable performance to the qutrit QNN, but utilized two qutrits, which may provide an avenue for exploring higher-dimensional quantum feature spaces. 
On the other hand, the qubit VQC fell behind, unable to achieve the same level of accuracy as its quantum counterparts. The classical SVM consistently outperformed all quantum methods, affirming the superiority of classical ML models in this particular scenario. These findings, although are expected to some extent, underscore the necessity for continued efforts to challenge classical ML models.

\section{Conclusions}

Several key observations emerge from the evaluation of the Gell-Mann feature map employed in the QKSVM and as an encoding layer for the QNN. The obtained results demonstrate remarkable promise, showcasing the feasibility of training by incorporating this encoding. It effectively expands the capacity and potential of quantum circuits.

However, it is essential to interpret these results with caution as they are based on simulations rather than physical hardware implementation, where noise significantly affects performance. Additionally, although the feasibility of training is demonstrated using the Gell-Mann feature map, classical methods continue to outperform their quantum counterparts. 

Furthermore, the effectiveness of the proposed methods heavily relies on the problem space and the underlying structure of the dataset. As heuristic methods, the selection of a different feature map or variations in the weights and layers of the variational model can entirely reshape the loss landscape.

Now, when comparing the QNN, a parameterized circuit with multiple layers, to the QKSVM, a notable observation emerged. It became evident that unless the variational circuit has a parameter count lower than the dimensions of the training data, kernel methods were more efficient. While the QNN exhibited slightly better accuracy in the examined problems, the improvement was marginal in most cases. Consequently, careful consideration should be given to this aspect.

Another intriguing distinction observed in the QKSVM method was that encoding the complete features of the dataset into the feature map using stacked layers yielded superior results compared to utilizing PCA and encoding the resulting principal components. Interestingly, the opposite trend was observed for the QNN. Several factors may contribute to this phenomenon.
This discrepancy can potentially be attributed to the increase in complexity and operations. As observed in the multiclass classification examples, surpassing a certain threshold of network layers rendered the model excessively challenging to train effectively. A similar observation holds true when increasing the depth of the encoding layer in the quantum kernel method.

Furthermore, an important observation from the  findings in simulations was that utilizing a single layer in most cases posed a significant obstacle during the training process for the QNN. The model encountered difficulty in surpassing the 50\% accuracy threshold, underscoring the inherent impracticability of capturing the requisite complexity of the data with just a single encoding and variational layer.
On the other hand, increasing the number of layers not only prolonged the training time for each epoch, but also extended the overall duration until the model converged. This outcome is expected, as introducing additional layers entails a higher number of parameters, requiring increased computational resources and optimization time.
However, despite the increased training time, the inclusion of more layers in the QNN architecture presented an opportunity to capture richer and more intricate representations of the data, leading to improved model performance. This trade-off between training time and enhanced expressiveness necessitates careful consideration when designing the QNN architecture.

Moreover, to overcome the challenges encountered when increasing the number of layers, exploring alternative network architectures could yield fruitful results. Another avenue worth exploring is the utilization of transfer learning techniques to enhance the initial stages of training, by initializing the model's weights with values derived from a pre-trained model. Considering that the weights often resided within similar ranges throughout the experiments, this approach can be particularly valuable when working with limited data, as it allows the model to leverage the data used in pre-training.

Finally, addressing challenges such as barren plateaus necessitated exploring strategies like structured initial guesses, local cost functions, and integrating correlations between layers. These strategies demonstrated effectiveness in the examined cases, but their heuristic nature requires further investigation. One promising avenue for optimization, that wasn't explored, is the utilization of Quantum Natural Gradient \cite{Stokes_2020} that leverages the Fubini-Study metric tensor to construct a quantum analog of natural gradient descent.

Looking ahead, future research in this area should prioritize the establishment of rigorous proofs, theoretical frameworks, and mathematical conditions for trainability. The demonstrated feasibility of qutrit quantum models using the Gell-Mann encoding underscores the promise of these methods. By developing a robust foundation that guides the optimization of these algorithms, we can fully harness the power of quantum computing to effectively address complex classification problems. Ultimately, this advancement will contribute to bridging the gap between classical and quantum ML approaches.


\acknowledgments 
AM and DS acknowledge partial support by the European Union’s Horizon Europe research and innovation program under grant agreement No.$101092766$ (ALLEGRO Project).

\bibliography{main2} 

\begin{thebibliography}{10}

\bibitem{Biamonte_2017}
Biamonte, J., Wittek, P., Pancotti, N., Rebentrost, P., Wiebe, N., and Lloyd, S., ``Quantum machine learning,'' {\em Nature}~{\bf 549},  195–202 (2017).

\bibitem{Petru}
Schuld, M. and Petruccione, F.,  [{\em Machine Learning with Quantum Computers}{\nolinebreak\hspace{0.1em}]}, Springer International Publishing (2021).

\bibitem{Gokhale_2019}
Gokhale, P., Baker, J.~M., Duckering, C., Brown, N.~C., Brown, K.~R., and Chong, F.~T., ``Asymptotic improvements to quantum circuits via qutrits,'' in [{\em Proceedings of the 46th International Symposium on Computer Architecture}{\nolinebreak\hspace{0.1em}]},  {ACM} (2019).

\bibitem{Blok_2021}
Blok, M.~S., Ramasesh, V.~V., Schuster, T., O'Brien, K., Kreikebaum, J., Dahlen, D., Morvan, A., Yoshida, B., Yao, N.~Y., and Siddiqi, I., ``Quantum information scrambling on a superconducting qutrit processor,'' {\em Physical Review X}~{\bf 11},  021010 (2021).

\bibitem{PMID:28658228}
Kues, M., Reimer, C., Roztocki, P., Cortés, L.~R., Sciara, S., Wetzel, B., Zhang, Y., Cino, A., Chu, S.~T., Little, B.~E., Moss, D.~J., Caspani, L., Azaña, J., and Morandotti, R., ``On-chip generation of high-dimensional entangled quantum states and their coherent control,'' {\em Nature}~{\bf 546},  622—626 (2017).

\bibitem{Wash_2023}
Wach, N.~L., Rudolph, M., Jendrzejewski, F., and et~al., ``Data re-uploading with a single qudit,'' {\em Quantum Machine Intelligence}~{\bf 5},  36 (2023).

\bibitem{mandilara2023classification}
Mandilara, A., Dellen, B., Jaekel, U., Valtinos, T., and Syvridis, D., ``Classification of data with a qudit, a geometric approach.'' arXiv: 2307.14060 (2023).

\bibitem{Schuld_2019}
Schuld, M. and Killoran, N., ``Quantum machine learning in feature hilbert spaces,'' {\em Physical Review Letters}~{\bf 122},  040504 (2019).

\bibitem{Havlicek_2019}
Havlíček, V., Córcoles, A., Temme, K., and et~al., ``Supervised learning with quantum-enhanced feature spaces,'' {\em Nature}~{\bf 567},  209–212 (2019).

\bibitem{Schuld_2020}
Schuld, M., Bocharov, A., Svore, K.~M., and Wiebe, N., ``Circuit-centric quantum classifiers,'' {\em Physical Review A}~{\bf 101},  032308 (2020).

\bibitem{P_rez_Salinas_2020}
P{\'{e}}rez-Salinas, A., Cervera-Lierta, A., Gil-Fuster, E., and Latorre, J.~I., ``Data re-uploading for a universal quantum classifier,'' {\em Quantum}~{\bf 4},  226 (2020).

\bibitem{Schuld_2019_grad}
Schuld, M., Bergholm, V., Gogolin, C., Izaac, J., and Killoran, N., ``Evaluating analytic gradients on quantum hardware,'' {\em Physical Review A}~{\bf 99},  032331 (2019).

\bibitem{hinton_2012_lecture}
Hinton, G., ``{Neural Networks for Machine Learning}.'' CSC321, Lecture 6 2012 \url{http://www.cs.toronto.edu/~tijmen/csc321/slides/lecture\_slides\_lec6.pdf}.

\bibitem{etingof2022lie}
Etingof, P., ``Lie groups and {Lie} algebras.'' arXiv:2201.09397 (2022).

\bibitem{merker2010theory}
Merker, J., ``Theory of transformation groups, by {S. Lie} and {F. Engel} (vol. i, 1888). {Modern} presentation and english translation.'' arXiv:1003.3202 (2010).

\bibitem{PhysRevA.76.042319}
Koch, J., Yu, T.~M., Gambetta, J., Houck, A.~A., Schuster, D.~I., Majer, J., Blais, A., Devoret, M.~H., Girvin, S.~M., and Schoelkopf, R.~J., ``Charge-insensitive qubit design derived from the cooper pair box,'' {\em Phys. Rev. A}~{\bf 76},  042319 (2007).

\bibitem{beautifulplots}
von Hahn, T., ``{Beautiful Plots: The Decision Boundary}.'' 2021 \url{https://www.tvhahn.com/posts/beautiful-plots-decision-boundary}.

\bibitem{Barren}
McClean, J.~R., Boixo, S., Smelyanskiy, V.~N., and et~al., ``Barren plateaus in quantum neural network training landscapes,'' {\em Nature Communications}~{\bf 76},  4812 (2018).

\bibitem{Cerezo_2021}
Cerezo, M., Sone, A., Volkoff, T., Cincio, L., and Coles, P.~J., ``Cost function dependent barren plateaus in shallow parametrized quantum circuits,'' {\em Nature Communications}~{\bf 12},  1791 (2021).

\bibitem{Grant_2019}
Grant, E., Wossnig, L., Ostaszewski, M., and Benedetti, M., ``An initialization strategy for addressing barren plateaus in parametrized quantum circuits,'' {\em Quantum}~{\bf 3},  214 (2019).

\bibitem{githublink}
Valtinos, T., ``{Quantum Neural Networks with Qutrits}.'' 2023 \url{https://github.com/Themiscodes/Quantum-Neural-Networks}.

\bibitem{Liu_2019}
Liu, J., Yuan, H., Lu, X.-M., and Wang, X., ``Quantum fisher information matrix and multiparameter estimation,'' {\em Journal of Physics A: Mathematical and Theoretical}~{\bf 53}(2),  023001 (2019).

\bibitem{lbfgs}
Liu, D.~C. and Nocedal, J., ``On the limited memory {BFGS} method for large scale optimization,'' {\em Mathematical Programming}~{\bf 45},  503--528 (1989).

\bibitem{cobyla}
Powell, M. J.~D.,  [{\em A Direct Search Optimization Method That Models the Objective and Constraint Functions by Linear Interpolation}{\nolinebreak\hspace{0.1em}]},  51--67, Springer Netherlands, Dordrecht (1994).

\bibitem{Stokes_2020}
Stokes, J., Izaac, J., Killoran, N., and Carleo, G., ``Quantum natural gradient,'' {\em Quantum}~{\bf 4},  269 (2020).

\end{thebibliography}
\bibliographystyle{spiebib} 
\end{document}